\documentclass[letterpaper]{aa} 
\usepackage{booktabs}
\usepackage{comment}
\usepackage{array}
\usepackage{graphicx}
\usepackage{txfonts}
\usepackage{natbib}
\usepackage[]{hyperref}

\newcommand{\unit}[1]{\mathrm{#1}}

\providecommand{\micron}{\ensuremath{\mu\mathrm{m}}}
\providecommand{\degree}{\ensuremath{^{\circ}}}

\newcommand{\inarcsecs}{\ensuremath{''}}

\begin{document}
  \title{Out-Of-Focus Holography at the Green Bank Telescope}
  \titlerunning{OOF Holography at the GBT}
  \authorrunning{B~Nikolic, et al}

  \author{B.~Nikolic\inst{1,2}, 
          R.~M.~Prestage\inst{1}, 
	  D.~S.~Balser\inst{1}, 
	  C.~J.~Chandler\inst{3} \and
          R.~E.~Hills\inst{4}}

  \institute{National Radio Astronomy Observatory, PO Box 2, Green
  Bank, WV 24944, USA \and 
  National Radio Astronomy Observatory, 520 Edgemont Road, 
  Charlottesville, Va 22903, USA (\email{bnikolic@nrao.edu}) \and 
  National Radio Astronomy Observatory, PO Box 0,  Socorro, NM 87801, USA \and
  Mullard Radio Astronomy Observatory, Cavendish Laboratory, Cambridge
  CB3 0HE, UK 
  }

  \hyphenation{holo-graphy}

  \abstract{We describe phase-retrieval holography measurements of the
  100-m diameter Green Bank Telescope using astronomical sources and
  an astronomical receiver operating at a wavelength of 7\,mm. We use
  the technique with parameterization of the aperture in terms of
  Zernike polynomials and employing a large defocus, as described by
  \cite{OOFNikolic05p1}. Individual measurements take around
  25\,minutes and from the resulting beam maps (which have peak signal
  to noise ratios of 200:1) we show that it is possible to produce
  low-resolution maps of the wavefront errors with accuracy around
  $\lambda/100$.
  
  Using such measurements over a wide range of elevations, we have
  calculated a model for the wavefront-errors due to the uncompensated
  gravitational deformation of the telescope. This model produces a
  significant improvement at low elevations, where these errors are
  expected to be the largest; after applying the model, the aperture
  efficiency is largely independent of elevation. We have also
  demonstrated that the technique can be used to measure and largely
  correct for thermal deformations of the antenna, which often exceed
  the uncompensated gravitational deformations during daytime
  observing.

  We conclude that the aberrations induced by gravity and thermal
  effects are large-scale and the technique used here is particularly
  suitable for measuring such deformations in large millimetre wave
  radio telescopes.  \keywords{telescopes} 
  }

\maketitle

\section{Introduction}

The performance of large single-dish radio telescopes operating at
millimetre and sub-millimetre wavelengths is often limited by the
accuracy and alignment of the optical elements. These inaccuracies may
arise during manufacture, assembly and adjustment, or they could be
due to uncompensated gravitational deformation, or due to external
factors such as differential heating or cooling of the structure.  A
commonly used method to measure such deformations using microwaves is
known as ``phase-retrieval holography'' \citep[see
e.g.][]{Morris1985,Anderson1985}. In this approach, only the amplitude
of the antenna pattern is measured, and the phase of the aperture
field (reflecting the deviations in the surface, and mis-collimation)
is recovered by numerical processing.  In this paper we describe the
application of a new phase-retrieval holography technique, discussed
in detail by \cite{OOFNikolic05p1}, to measure and correct such
inaccuracies at the 100-metre diameter Robert C. Byrd Green Bank
Telescope \citep[the GBT, see e.g.,][]{2004SPIE.5489..312J}.

The GBT has a very capable active surface control system
\citep{1998SPIE.3351..310L}. The primary reflector is composed of 2004
panels mounted in rings that are concentric to the vertex of the
208\,m parent (virtual) parabola.  The surface accuracy of individual
panels is about 75\,\micron. The panels are mounted at their corners
on computer-controlled actuators such that the corners of four
adjacent panels share one actuator; the actuators can be positioned to
within a tolerance of 25\,\micron. The requirement is then to provide
appropriate commands to the active surface control system to correct
for the inaccuracies described above.  Our long-term goal is to
deliver a surface accuracy of $\sim$ 200\,\micron, suitable for
telescope operation up to 115\,GHz. Our intermediate goal, to allow
operation up to $\sim$ 50\,GHz, has a corresponding surface accuracy
requirement of 360\,\micron\ independent of elevation. Currently, the
active surface is operated open-loop, using a simple look-up table of
corrections as a function of elevation. Initially, these corrections
were derived from a finite element model of the antenna. Although this
provided a modest improvement over the static (photogrammetry)
zero-point correction, there remained a significant fall-off in
surface accuracy at low and high elevations. The need to provide an
improved look-up table to correct for the unmodelled gravitational
errors, and potentially also to correct for thermal effects, provided
the motivation for the work described in this paper.

By using a parametrisation scheme employing Zernike polynomials and by
making maps with a large defocus, the new measurement
technique---which we colloquially refer to as Out-Of-Focus (OOF)
holography---requires a significantly smaller dynamic range than
``traditional'' phase-retrieval holography measurements and hence can
make use of astronomical sources and receivers.  This in turn allows
measurements to be made at a wide range of telescope elevations.
Also, using the facility astronomical receivers means that each
measurement provides a map of wavefront errors for the complete system
used for astronomical observations, i.e., the combination of errors
arising from surface inaccuracies, any mis-collimation and the phase
response of the receiver.

The smaller required dynamic range is, however, achieved through a
trade-off in the possible resolution in the map of wave-front errors
(or, aberrations): typically we describe the wavefront errors as a
linear combination of 20--30 basis vectors. This means that OOF
holography measurements are only sensitive to wavefront errors which
are correlated over large scales in the aperture plane---in the case
of the GBT, these spatial scales are around 20\,m or larger.

This low resolution makes OOF holography unsuitable for panel-to-panel
adjustment of large radio telescopes and for assessing the quality of
individual panels.  The ability to make measurements at a wide range
of elevations, however, makes it possible to measure aberrations
induced by gravitational deformations
(Section~\ref{sec:grav-deform}). The ability to make measurements
(comparatively) quickly, using the standard astronomical receivers
with no special hardware setup, means that the technique can also
potentially be used to measure and correct for non-repeatable effects,
specifically thermal deformations, in ``real-time''.

\subsection{GBT Surface Setting and Classes of Deformations/Corrections}

The GBT is designed to operate over a wide range of wavelengths, from
approximately one meter (300\,MHz), through intermediate centimeter
wavelengths, in to the millimeter regime (3mm or 100 GHz). At the
longest wavelengths, the antenna optics are essentially perfect. A
number of successively more precise corrections are required to allow
the antenna to operate effectively at the shorter wavelengths.

Prior to this work, the first three types of corrections described below
were routinely being applied. Development of the fourth type of
correction is a direct outcome of the work described here.

\subsubsection{Construction misalignment correction at the "rigging angle"}

During the construction of the GBT, one retro-reflector was placed on
one of the panel corners adjacent to each actuator, and the locations
of all 2004 retro-reflectors were measured using
photogrammetry. Actuator zero-points were then derived which, as far
as possible given the accuracy of the photogrammetry, compensate for
actuator misalignments during the construction process. These settings
were chosen so that the primary surface would be correct at the
rigging angle of 50.3 degrees. The typical actuator zero-point
correction is 1.9\,mm, and the root-mean-square (RMS) panel to panel
surface error after this correction is expected (from the accuracy of
the photogrammetry) to be approximately 250\,\micron.

\subsubsection{Homologous deformation and subreflector "focus tracking"}

Like all modern large radio telescopes, the GBT is built to a
homologous design \citep{1967AJ.....72...35V}. As the antenna tips
from the zenith to the horizon, the primary mirror deforms due to the
effect of gravity, but in such a way that the antenna remains
(approximately) a parabola, and that the main effect of the
deformation is simply to change the focal length of the parabola. At
the same time, the offset feed-arm, which supports the sub-reflector
and the receiver cabin at the Gregorian focus, also deflects under
gravity. All of these effects are compensated for via a lookup table
of sub-reflector lateral and radial focus translations as a function
of elevation, which refocus the antenna to maximise the gain at the
Gregorian focus.

\subsubsection{Finite Element Model (FEM) correction for non-homologous deformation}

Due to the off-axis design of the GBT, it is not fully homologous. A
finite element (FE) model is used to predict the displacements of the
primary mirror backup structure as a function of elevation relative to
its rigging angle position.  These displacements are then used to
calculate both the "best-fit" paraboloid, and the residuals from this
paraboloid. These residuals are stored in a look-up table, and applied
to the antenna via the actuators as the telescope tips in
elevation. As an indication of the magnitude of this correction, the
RMS of the FE model actuator corrections at 15\degree\ elevation is
880\,\micron.

Together, the actuator zero-point corrections for construction
misalignments, the subreflector focus-tracking to accomodate the gross
deformations due to the homologous design, and the FE model derived
non-homology corrections allow operation of the antenna at wavelengths
as short as 1.5\,cm (frequencies up to 20\,GHz).

\subsubsection{Residual non-homology, thermal deformations and large-scale setting
error at the rigging angle}

After applying all of the above corrections, it was apparent from
observations at even shorter wavelengths (7\,mm, or 43\,GHz) that
there remained two further large-scale surface error contributions. 1)
Residual non-homology, or unmodelled gravitational errors. Despite
application of the FE model, the gain-elevation curve continued to
show a systematic and repeatable fall-off in gain to high and low
elevations, implying that the FE model was not accurately capturing
the residual gravitational deformations. (This is not surprising, as
we suspect from independent comparisons that the FE model is
systematically in error). The most likely reason for this error is
that the FE model has not been yet updated for changes to the
telescope structure and weight distribution during construction and
operations.  2) Thermal deformations. Observations during daytimes
demonstrated that thermal gradients could induce deformations of the
antenna, and a resulting loss of efficiency, which could easily exceed
the loss of efficiency due to un-modeled gravitational deformations.

Finally, and not obvious from efficiency measurements alone, our OOF
investigations revealed that even at the rigging angle under benign
thermal conditions, there appeared to be a static, large-scale surface
error.

All three of these error components are amenable to investigation, and
correction, via the OOF technique.

\section{Measurement Strategy}

For all of the measurements presented here, the data consisted of
three (total-power) beam-maps taken consecutively, the first with the
telescope in-focus and the other two with the telescope $\pm
35\,\unit{mm}$ (five wavelengths) out of focus. The de-focusing of the
telescope was carried out by moving the sub-reflector in its ``Y''
direction, that is, in the direction along which its motion does not
change the centroid of the beam on the sky. The convention used at the
GBT is that a positive displacement corresponds to moving the
sub-reflector away from the primary reflector.

The receiver used for these observations was the facility Q-band
receiver operating at 43.1\,GHz and the bandwidth employed was either
80\,MHz or 320\,MHz, depending on the date of observation.
This receiver consists of two feeds, each with two polarizations, and
thus has two beams on the sky, separated by 58 arcseconds ($\sim3.5$
FWHM) in the azimuth direction. To minimise the effect of
sky-brightness variations on the measured beam-maps, the output of
these two receivers was differenced so that the effective response of
the telescope was the real beam convolved by two delta functions
separated by 58 arcseconds in the azimuth direction. The aberrations
due to both of the feeds being off (and on opposite sides of) the
optical axis are negligible in this instance and were not taken into
account in the further analysis.

All observations presented here were carried out in continuum
mode. The sources used were chosen to be strong (all were stronger
than 5\,Jy) and compact with respect to the nominal GBT beam at
43\,GHz (16 arcseconds); we mainly used the sources J0319+4130 (3C84),
J1256-0547 (3C279), J2148+0657, and J2253+1608 (3C454.3). Since the
system temperature at Q-band is typically around 75\,K, the
theoretical signal to noise ratios expected are in the region of
$500:1$.  Due to gain and atmospheric fluctuations, the actual peak
signal to noise ratio we achieved was of the order of $200:1$. We have
also made successful out-of-focus beam maps using spectral-line
observations of strong SiO masers, although none of these data are
used in the analysis presented later.

The beam maps were obtained using the on-the-fly technique (in the
Alt/Az frame), that is, the telescope was continuously in motion and a
high sampling rate (10\,Hz) used to avoid excessive smearing in the
in-scan direction. The maps consisted of seventeen azimuth rows each
350 arcseconds in length and separated by eight arcseconds in the
elevation direction. The scanning speed of the telescope was fifty
arcseconds per second of time and ten seconds was allowed for turn
around at the end of each row; hence, the total time required for each
map was just under five minutes and the integration time per Nyquist
sample was 0.3 seconds. The entire trajectory was optimised to avoid
jerking the telescope during end-of-row turn around. Additionally, the
long length of the rows relative to the beam size allows time for any
oscillations of the telescope to damp down before it reaches the
source. A sample set of observed beam maps (re-gridded for display
purposes) is shown in the top row of Figure~\ref{fig:samplebeam}.

In practice each set of three beam-maps was preceded by pointing and
focus measurements although this is not strictly necessary.

\section{Analysis strategy}

\begin{figure*}
  \begin{centering}

  \includegraphics[angle=-90,width=0.31\linewidth]{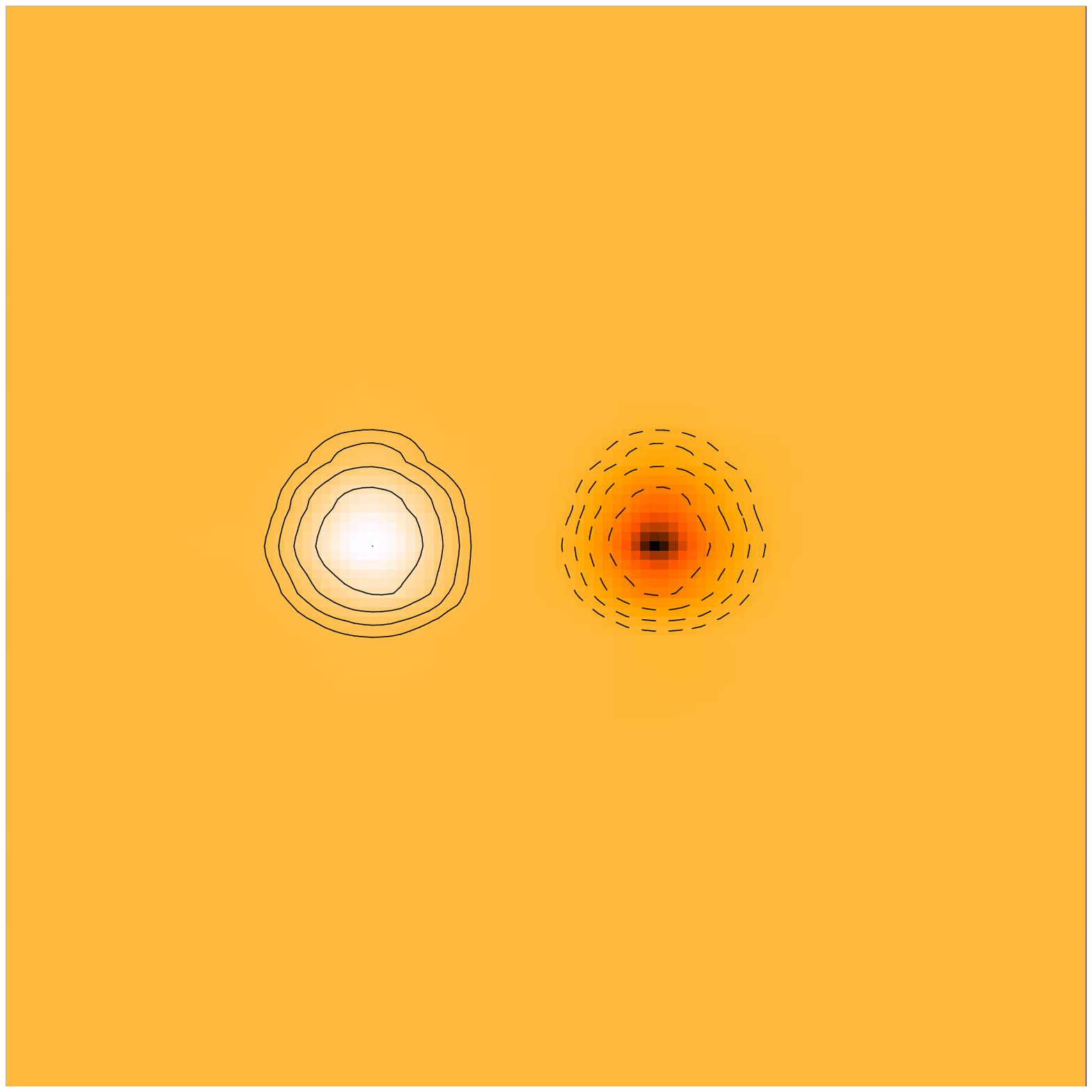}
  \includegraphics[angle=-90,width=0.31\linewidth]{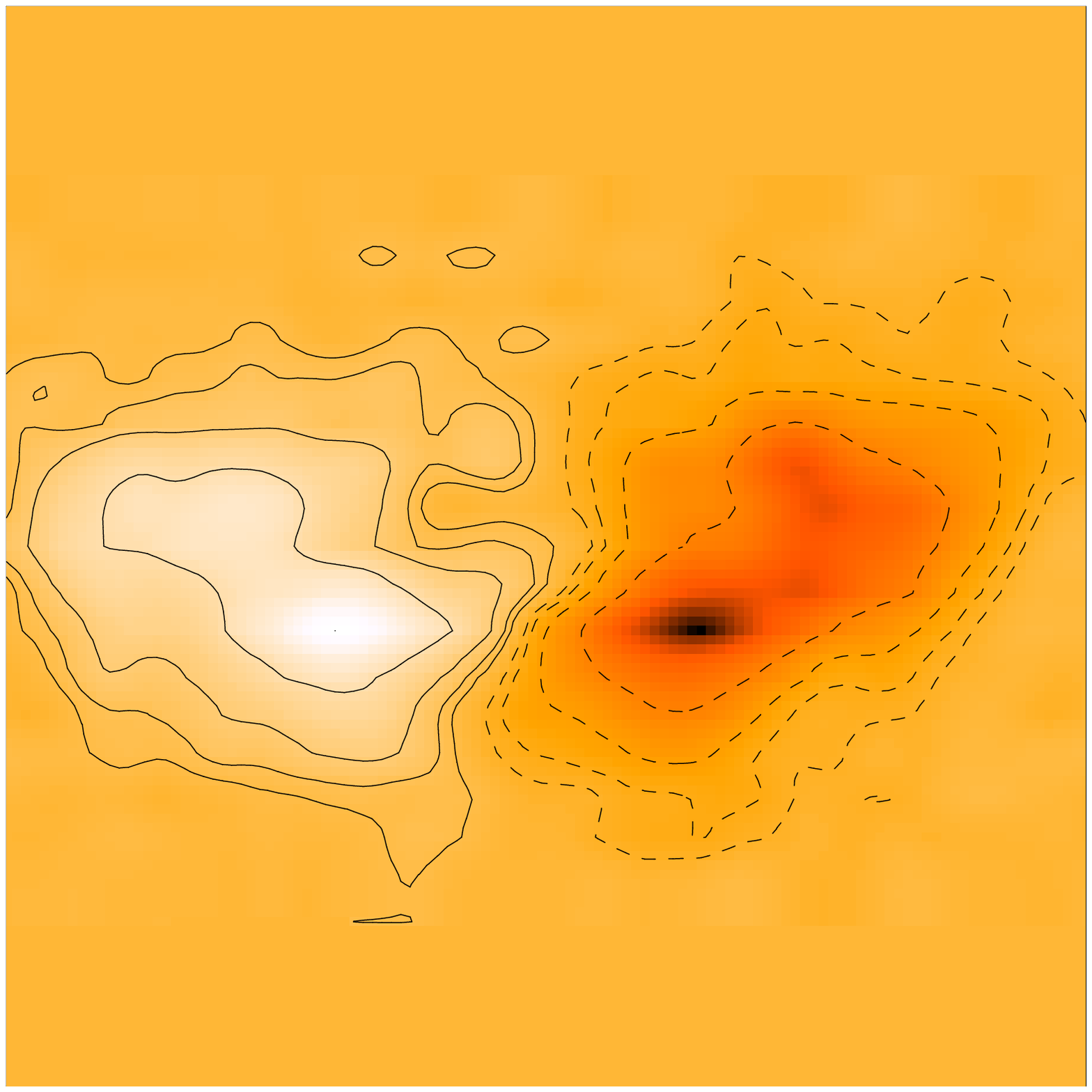}
  \includegraphics[angle=-90,width=0.31\linewidth]{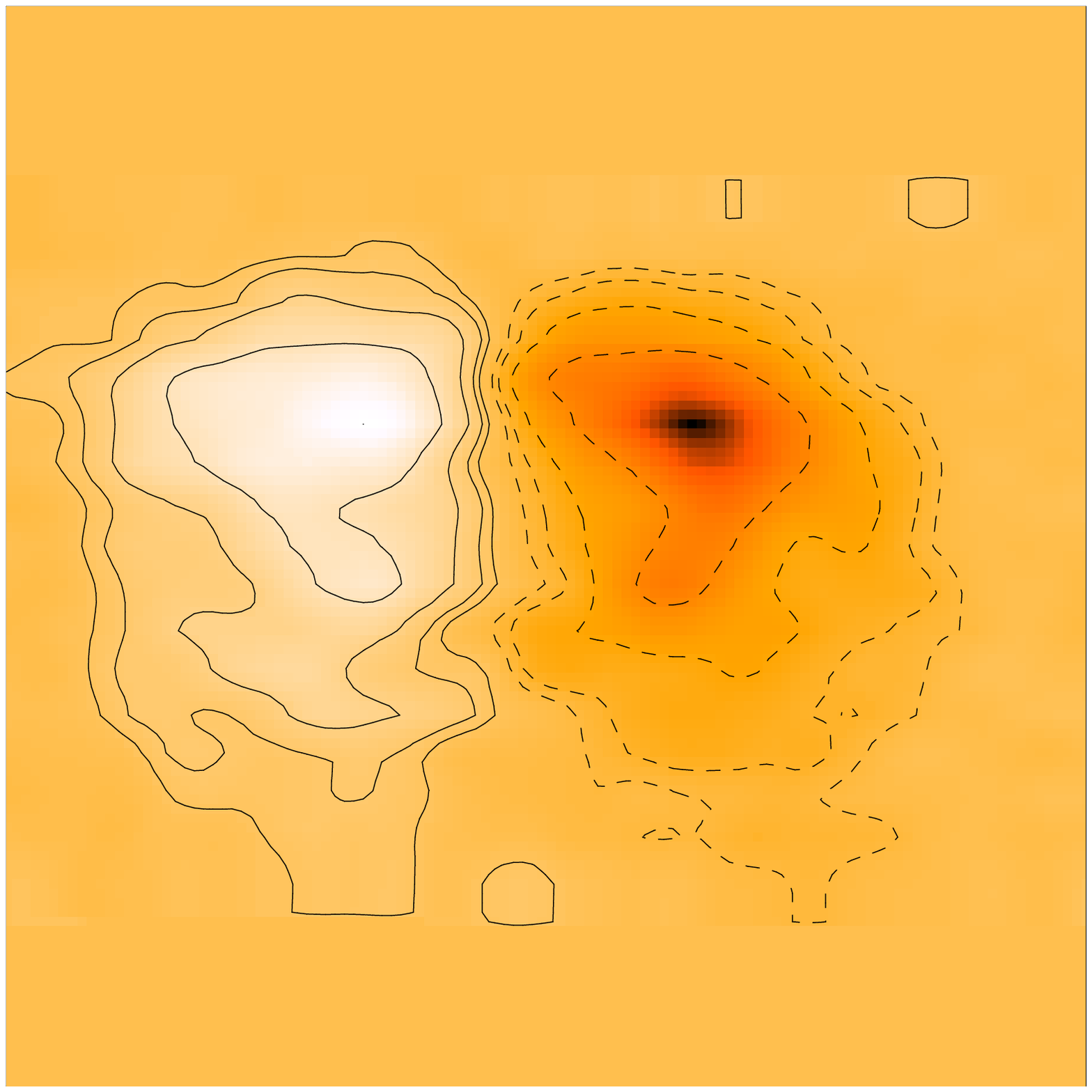}

  \includegraphics[angle=-90,width=0.31\linewidth]{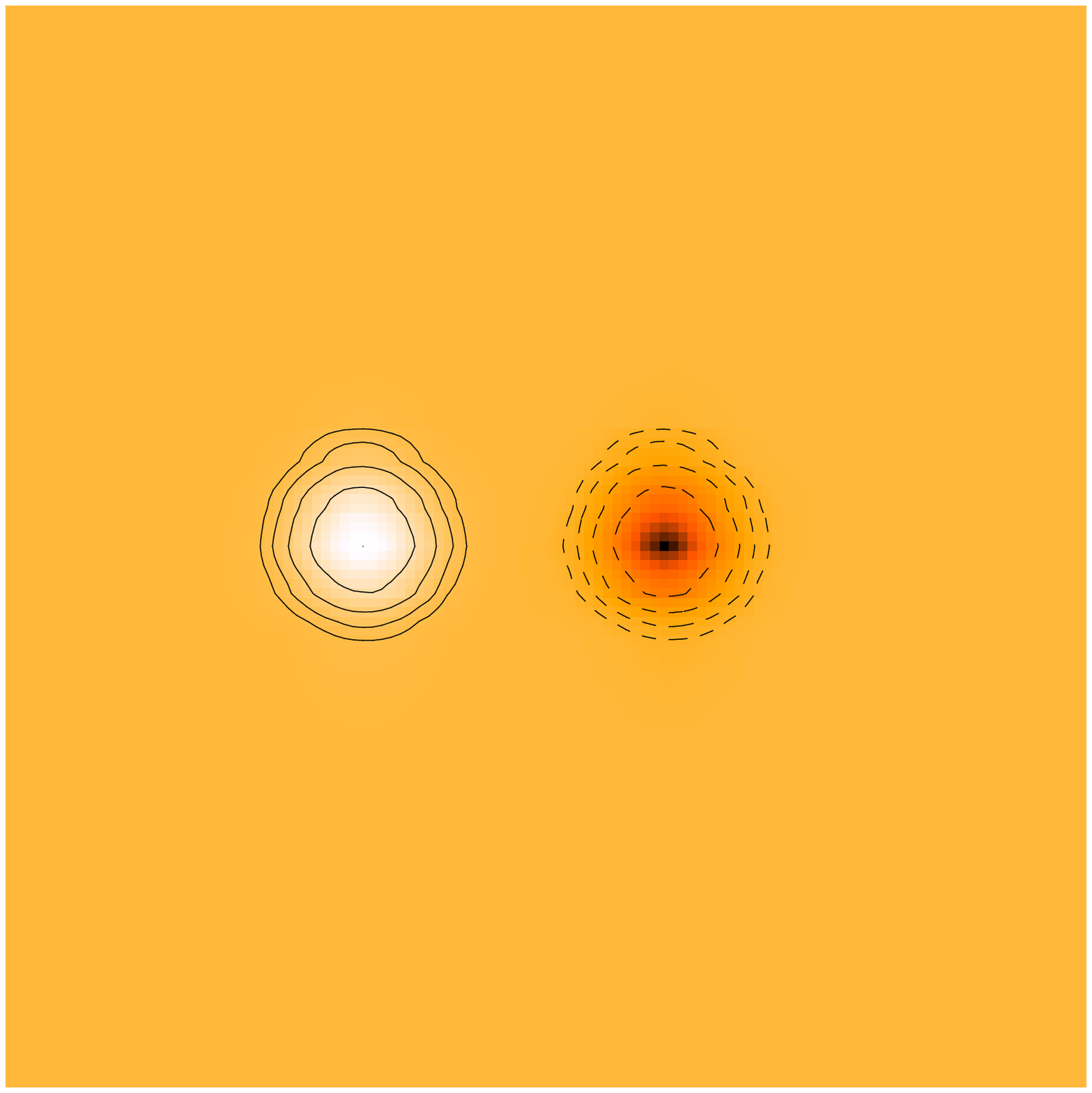}
  \includegraphics[angle=-90,width=0.31\linewidth]{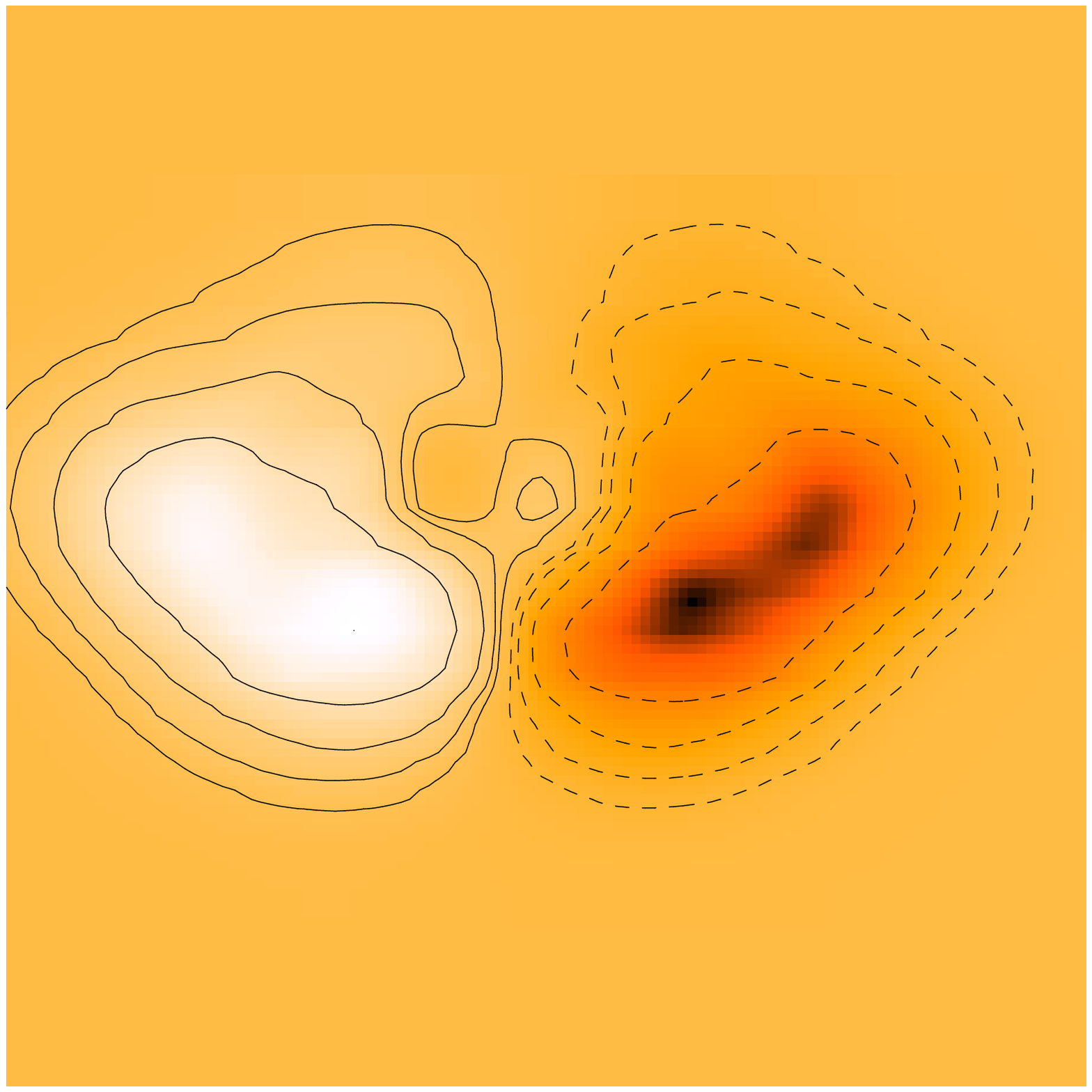}
  \includegraphics[angle=-90,width=0.31\linewidth]{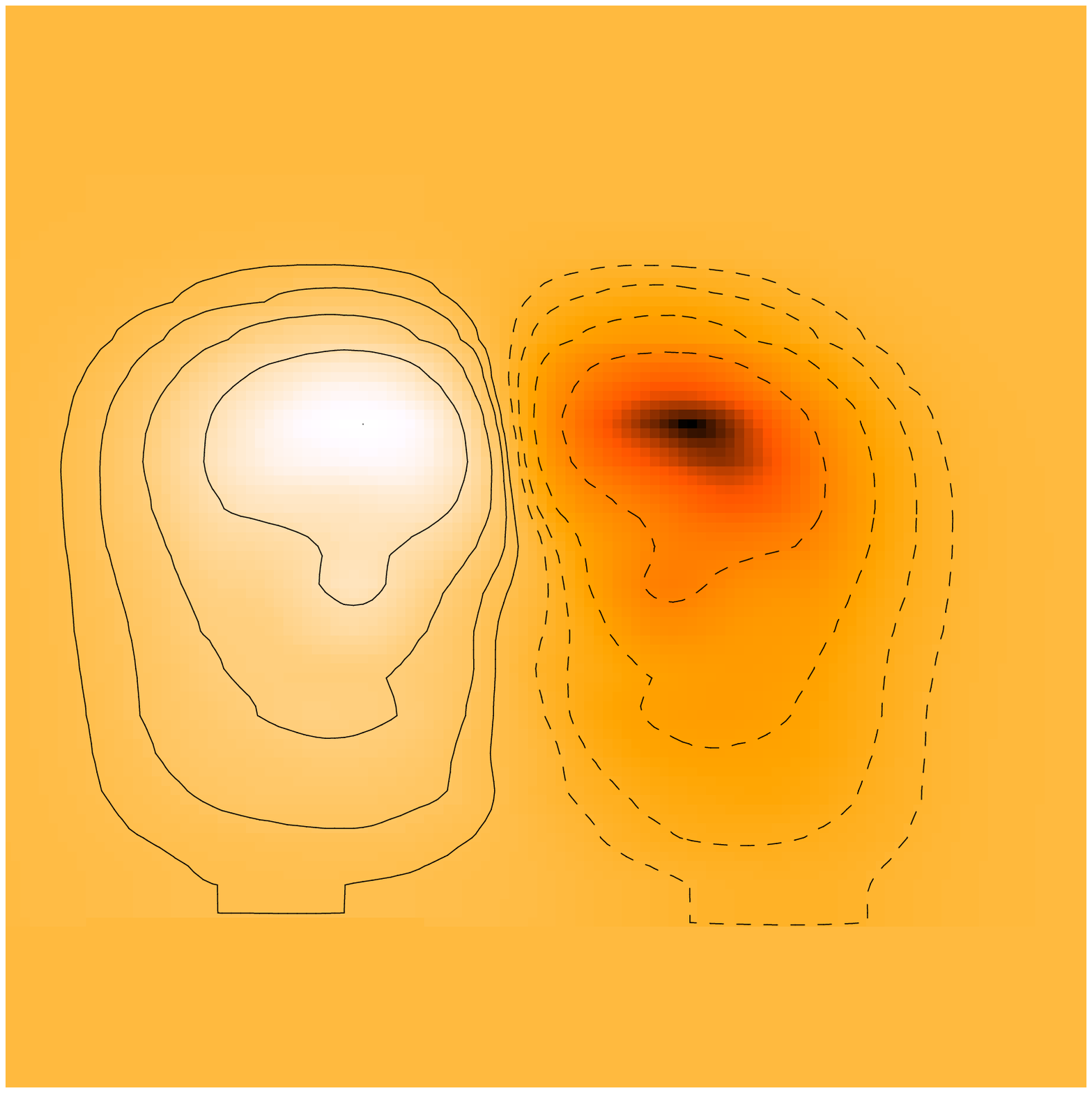}

  \end{centering}

  \caption{Top row: A sample set of 43\,GHz out-of-focus maps taken on
  11th April 2005 using 3C279 as the source. The left-most map is
  in-focus, the other two are both $5\, \lambda$ out of focus, with
  the sense of the defocus such that in the centre map the
  subreflector is further away from the primary than the optimal focus
  and in the right-most map the subreflector is closer to the primary
  than optimal. Bottom row: simulated beam maps of the best-fitting
  model (using Zernike polynomials up to fifth radial order inclusive)
  to the observed maps in the top row.  All of the maps show an area
  of $210\inarcsecs \times 210\inarcsecs$ and have the elevation
  direction running vertically and azimuth direction horizontally. The
  contours are drawn at logarithmic intervals.}

  \label{fig:samplebeam}

\end{figure*}

The raw data recorded by the telescope needed some minor processing
before being analysed with the OOF software. This consisted
of calibrating (using the noise-injection diodes), differencing the
signals from the two feeds, and removing residual sky brightness
variations by fitting linear baselines to the ends of each row and
subtracting these baselines from the recorded data.

The resulting time series data consists of antenna offsets (relative
to the position of the source) in the azimuth and elevation directions
and the antenna temperature for each sample. The data for a set of
three out-of-focus beam maps were then analysed using the custom OOF
software to produce a map of the total wavefront errors present in the
telescope. For the purposes of OOF analysis it is neither necessary
nor desirable to regrid these data onto a regular grid. The reason for
this is that the \emph{model} beams can just as well be interpolated
to the observed points while regriding the observed data will
inevitably cause some information loss.

The technique for analysing beam maps to infer the aberrations is
described in detail by \cite{OOFNikolic05p1}. In that paper we
considered a classical Cassegrain telescope design.  The GBT has an
off-axis Gregorian design, and so the phase change across the aperture
plane induced by moving the secondary reflector is different from the
Cassegrain case, and in fact not circularly symmetric, as can be seen
from the asymmetry in the elevation direction of the observed beam
maps in Figure~\ref{fig:samplebeam}.  But, as long as the correct
expression for the phase change due to defocus is used when
calculating the model beams, this difference does not affect the
performance of the OOF technique in any way.

In the present study we parametrise the possible aberrations as a
linear combination of the first five radial orders of Zernike's circle
polynomials \citep[see e.g.,][]{BORNWOLF}, which we label as:
\begin{equation}
  Z_{n,l}(r, \theta)= \left\{ \begin{array}{ll}
      R_{n,l}(r) * cos( l \theta) &
      l \geq 0 \\
      R_{n,l}(r) * sin( - l \theta) &
      l < 0.
      \end{array} \right.
\end{equation}
Here, $R_{n,l}(r)$ is the radial dependence of the Zernike polynomial
as given by \cite{BORNWOLF}and $\theta$ is the angle measured from the
cross-elevation (left-right in the plots below) direction .  Since
there are $(n+1)(n+2)/2$ polynomials up to and including radial order
$n$, this gives a total of 18 non-trivial polynomials (the polynomials
corresponding to piston and tilts being considered trivial in this
instance) which form the basis vectors for the space searched by the
optimisation algorithm. The aperture plane amplitude was described as
a well-centred, circular Gaussian with a width corresponding to an
edge taper of 14.5\,dB, close to the design value of the receiver we
used. Our initial investigations did not show evidence for significant
deviation of these illumination parameters so they were fixed at the
above values when fitting for the aberrations shown below.  There were
only two further free parameters: the overall amplitude of the
illumination and the relative gains of the two feeds.

In summary, the free parameters supplied to the minimisation procedure
were the 18 non-trivial coefficients of Zernike polynomials, the two
tilt terms (to account for the possibility of an error in the
telescope pointing), the relative gains of the beams and the overall
amplitude.

\section{Validation, Repeatability and Closure}

We have performed extensive tests to confirm the accuracy and
repeatability of this measurement technique. Firstly, we have
deliberately introduced a known aberration into the telescope optics,
and confirmed that we can correctly measure this. Secondly, we have
performed a number of repeat OOF measurements, applying the results of
the first before performing the second, and confirming that the
surface has actually improved. Finally, we have confirmed the success
of the applied correction by independent astronomical measurements.
Each of these is discussed in detail below.

\subsection{Retrieval of a Known Deformation}
\label{sec:retr-known-deform}

In this experiment, we used the active surface system of the GBT to
introduce a known ``bump'' into the figure of the primary
reflector. The peak of the bump was 770\,\micron. The bump was
decomposed into a linear combination of the first five orders of
Zernike polynomials and applied by specifying the coefficients to the
active surface control system.  The procedure for applying the
coefficients is exactly the same as that which is used to make
corrections to the surface from results of OOF holography, and so this
experiment, in addition to testing the ability of the OOF technique to
measure aberrations, also provided a useful end-to-end test of the
system, including coordinate system conventions, Zernike coefficient
ordering and so on.

\begin{figure}
\begin{centering}

  \includegraphics[width=0.8\columnwidth,angle=-90]{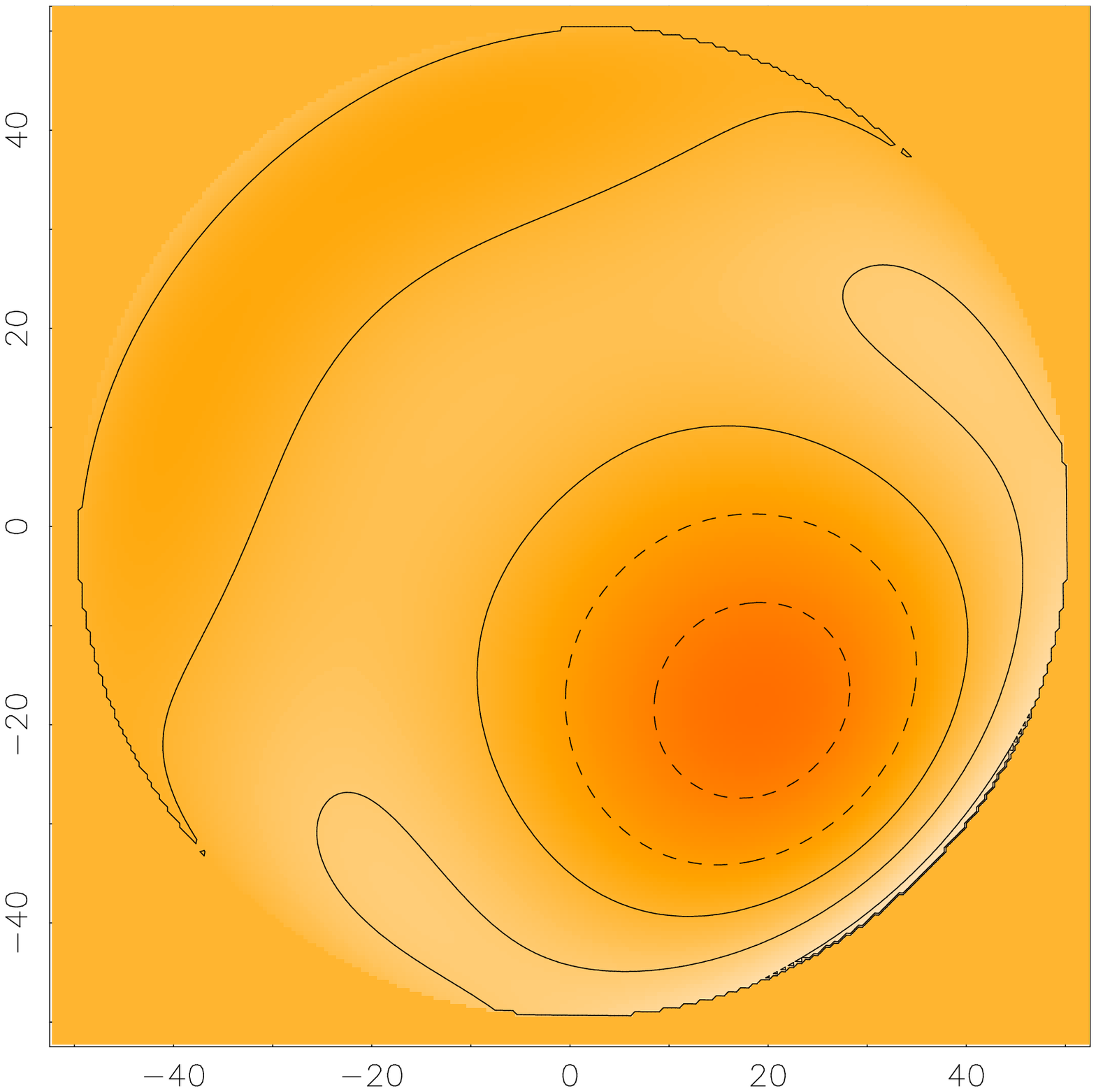}

  \includegraphics[width=0.8\columnwidth,angle=-90]{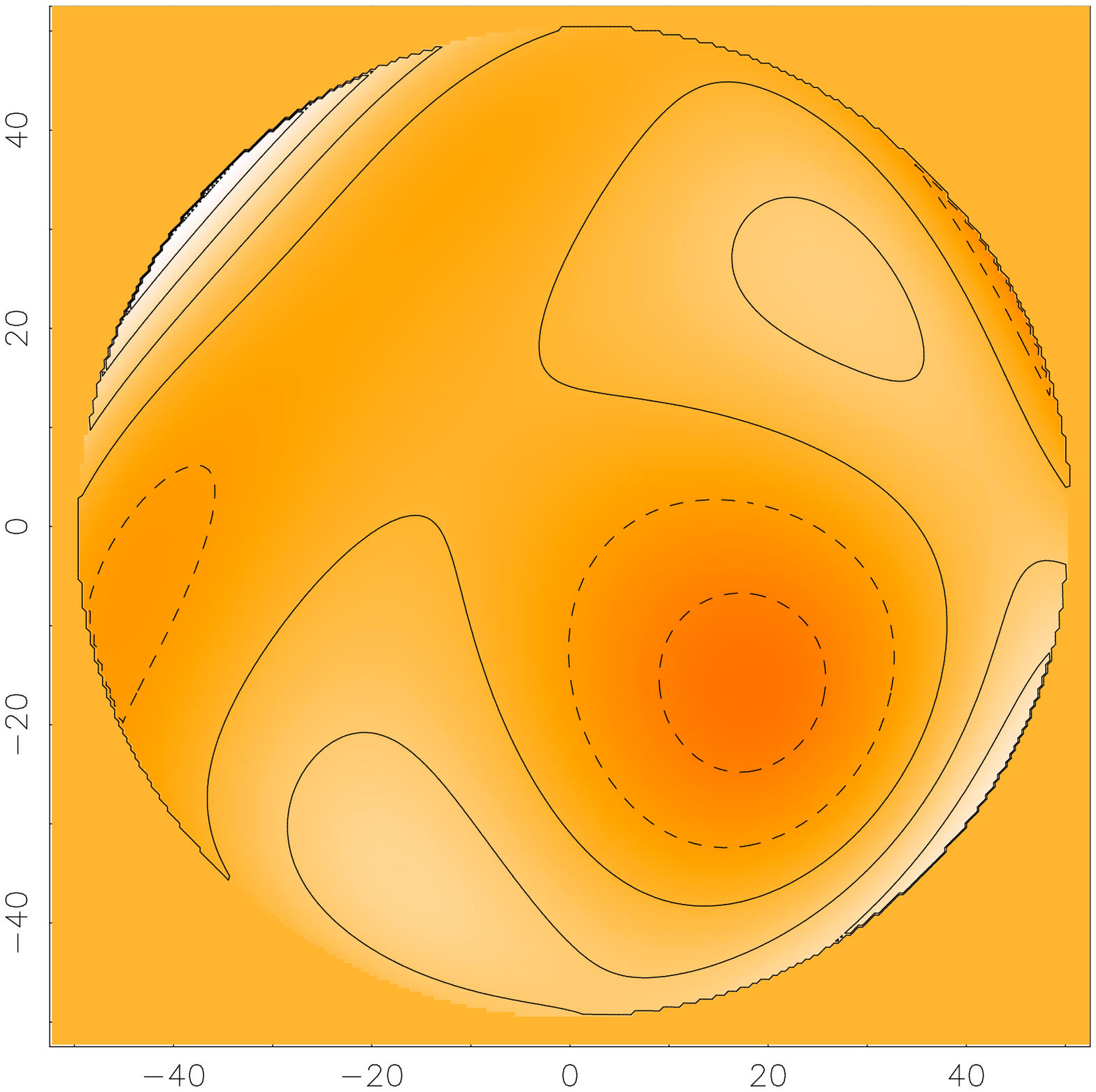}

  \caption{Test of the OOF technique by measurement of a known
  wavefront error.  Top panel: wavefront error introduced by changing
  the figure of the primary reflector. Bottom panel: estimate of this
  wavefront error using the OOF holography technique. The wavefront
  error is in radians of phase (one radian corresponds to a surface
  displacement 550\,\micron\ in the normal direction) and contours are
  drawn at half-radian intervals. Grayscale is
  $\pm2\,\unit{radians}$.}

  \label{fig:gbtbump}

\end{centering}
\end{figure}

The test consisted of three sets of OOF measurements: the measurement
with the bump applied to the surface; and two measurements, proceeding
and following the bump measurement, with the nominal surface. The mean
of these nominal-surface measurements was used to estimate and remove
aberrations not due to the introduced bump.  The results are shown in
Figure~\ref{fig:gbtbump}. The top panel shows the wavefront error
expected from the bump introduced onto the surface. The bottom panel
shows the OOF measurement of wavefront error introduced by the bump,
i.e., it is the aperture phase distribution retrieved from beam maps
with the bump applied minus the mean aperture phase distribution
retrieved from measurements without the bump.

As can be seen from the figure, the measured and expected phase
distributions are in fair agreement. As is typical for OOF holography
measurements, the largest errors in the retrieved phase distribution
are at edges of the dish, which are less well constrained due to the
tapered illumination pattern of the astronomical receiver. The
illumination-weighted half-path RMS difference between the bump and
no-bump measurements -- i.e., the value which can be used in the
\cite{RuzeEff} formula to calculate the expected decrease in aperture
efficiency due to residual large-scale structure -- is 130\,\micron;
this error is due to the combination of the intrinsic error of the
technique and the error in the estimate of aberrations present with no
bump applied to the surface.

\subsection{Repeatability, Closure and comparison with Independent 
Measurements}
\label{sec:closure}

Another way to investigate the practical accuracy of the OOF
holography technique is to measure the best closure that can be
achieved using the technique. This involves observing a set of OOF
beam maps and analysing them to obtain the aberrations, compensating
for these aberrations by adjusting the primary surface and then
repeating the beam map measurement to evaluate how well the initially
observed aberrations were compensated for. Because the GBT primary
surface is ``active'', it can be easily and essentially instantaneously
adjusted, making it possible to observe both of the sets of beam maps
within about one hour.

There are some obvious sources of error with this approach, most
notably that the actual aberrations are likely to be changing
continuously with time and so correcting for the aberrations at the
time of the first measurement can not be expected to fully remove the
aberrations at the time of the validation measurement. Nevertheless, a
measurement of the closure does provide a good upper limit on the
\emph{random} errors associated with the OOF technique. As this is
essentially a self-consistency measurement, it provides no constraint
on systematic errors that may be present in the technique. However,
independent measurements (such as for example a measurement of the
aperture-efficiency of the telescope) can confirm whether the expected
performance improvement has been realised.

\begin{figure}
\begin{centering}

  \includegraphics[width=0.8\columnwidth,angle=-90]{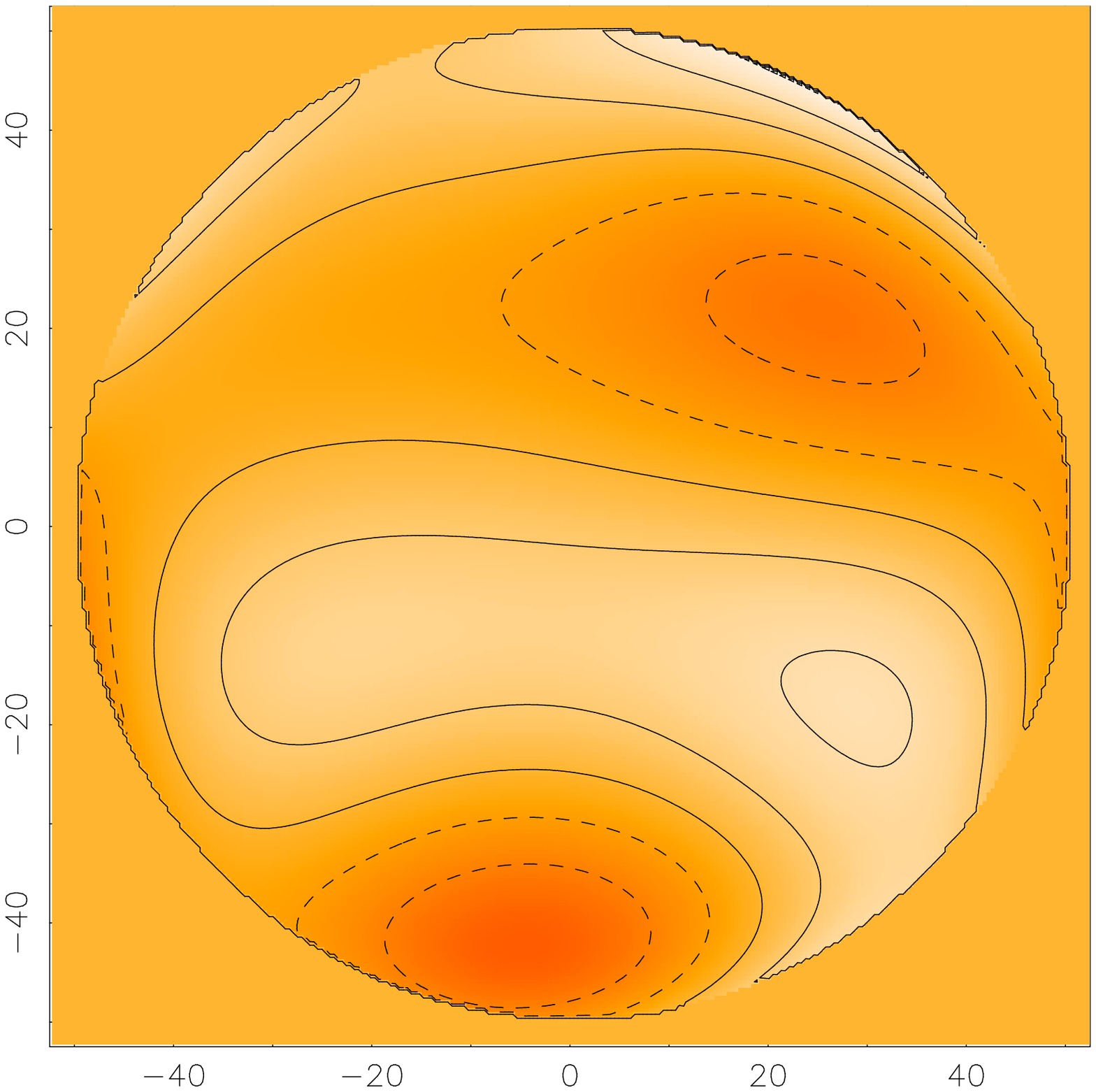}

  \includegraphics[width=0.8\columnwidth,angle=-90]{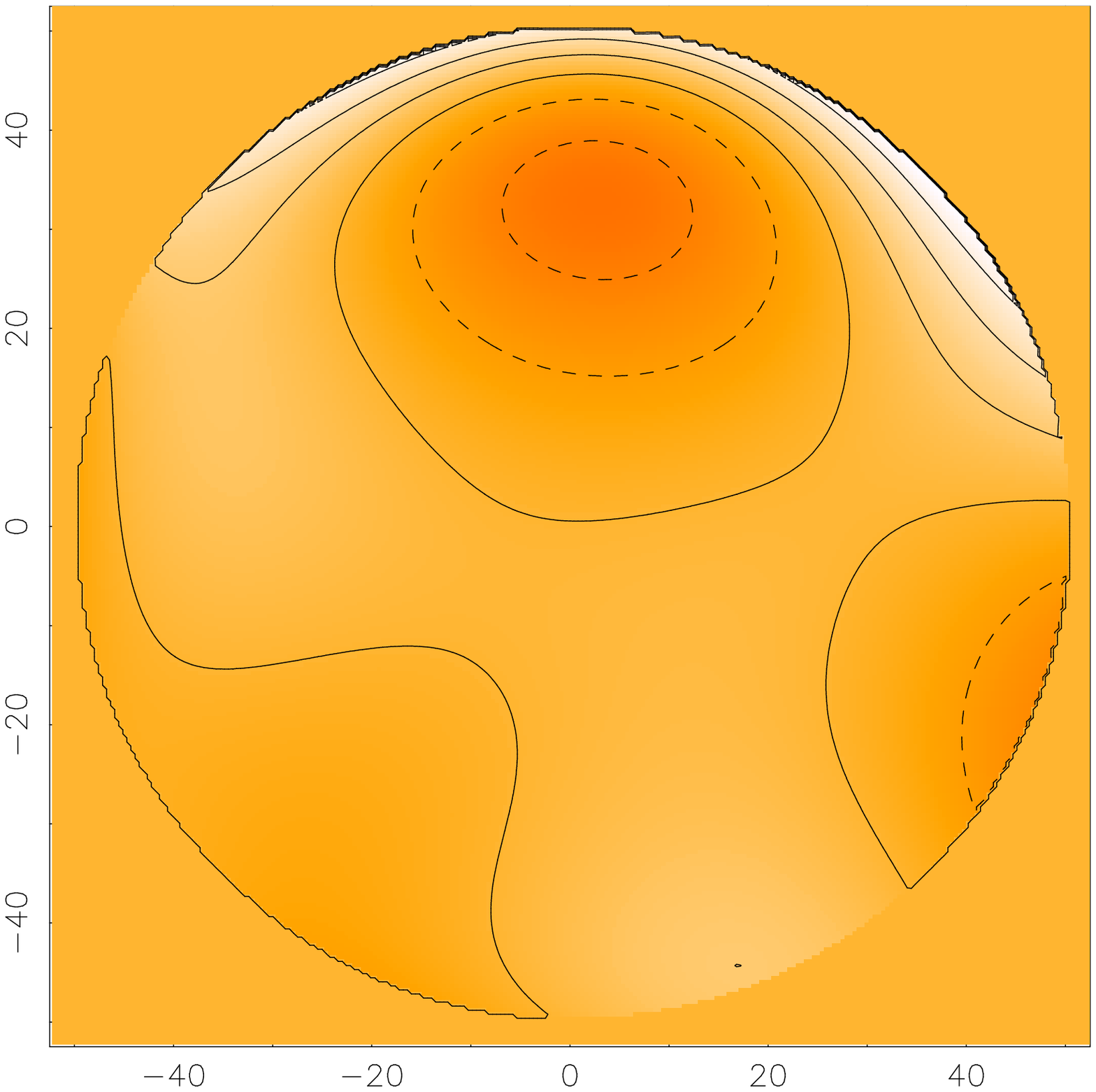}

  \caption{Closure of the OOF technique at the GBT. 
    Top: wavefront errors derived from daytime OOF measurements; the
    corresponding illumination-weighted half-path surface RMS is
    330\,\micron.
    Bottom: the measured wavefront error distribution after applying
    corrections derived from the measurement shown in the top
    panel, obtained one hour earlier. The weighted RMS is reduced 
    to 220\,\micron.
    Contours in both maps are at half-radian intervals and the
    grayscale range is $\pm2\,\unit{radians}$.  }

  \label{fig:gbtclosureday}

\end{centering}
\end{figure}

We have performed a number of measurements of this type.  A
particularly dramatic example is shown in
Figure~\ref{fig:gbtclosureday}.  In this case, the initial measurement
was made at 11\,am local time on 11th April 2005, a bright sunny day.
The initially measured large scale aberrations (top panel of
Figure~\ref{fig:gbtclosureday}) were, because of the time of day and
resulting thermal deformations, rather large; they correspond to an
illumination-weighted, half-path RMS wavefront error of around
330\,\micron.  These OOF measurements were processed and applied to
the telescope approximately an hour later, and another set of OOF maps
with these corrections applied was taken. The bottom panel of
Figure~\ref{fig:gbtclosureday} shows the aberrations derived from this
second set of OOF maps, that is the residual aberrations after
correcting for what we measured in the top panel. The estimated
weighted half-path RMS derived from this second set of OOF maps is
220\,\micron.

This particular closure experiment was not ideal due to the
significant time delay between the two measurements, which means that
the actual thermal and gravitational deformations of the telescope
would have changed between them.  However, it was chosen to illustrate
both the real improvement that this technique can provide, and that it
can be used to correct wavefront errors which are clearly dominated by
thermal deformations.

Under more ideal (benign night-time) conditions, with the second
measurement made within 30 minutes of the first, typical half-path RMS
residuals of ~100\,\micron\ or less were obtained.  This residual
reflects the random error associated with two OOF measurements, and so
is a factor of $\sqrt 2$ larger than the error of single measurement,
which is therefore around 70\,\micron\ illumination-weighted half-path
RMS when operating at a wavelength of 7\,mm and using a basis set
consisting of the first five radial orders of Zernike polynomials. In
terms of the observing wavelength, the accuracy of the technique (at
the scales being probed) is around $\lambda/100$.

\begin{figure}
\begin{centering}

  \includegraphics[width=0.99\columnwidth]{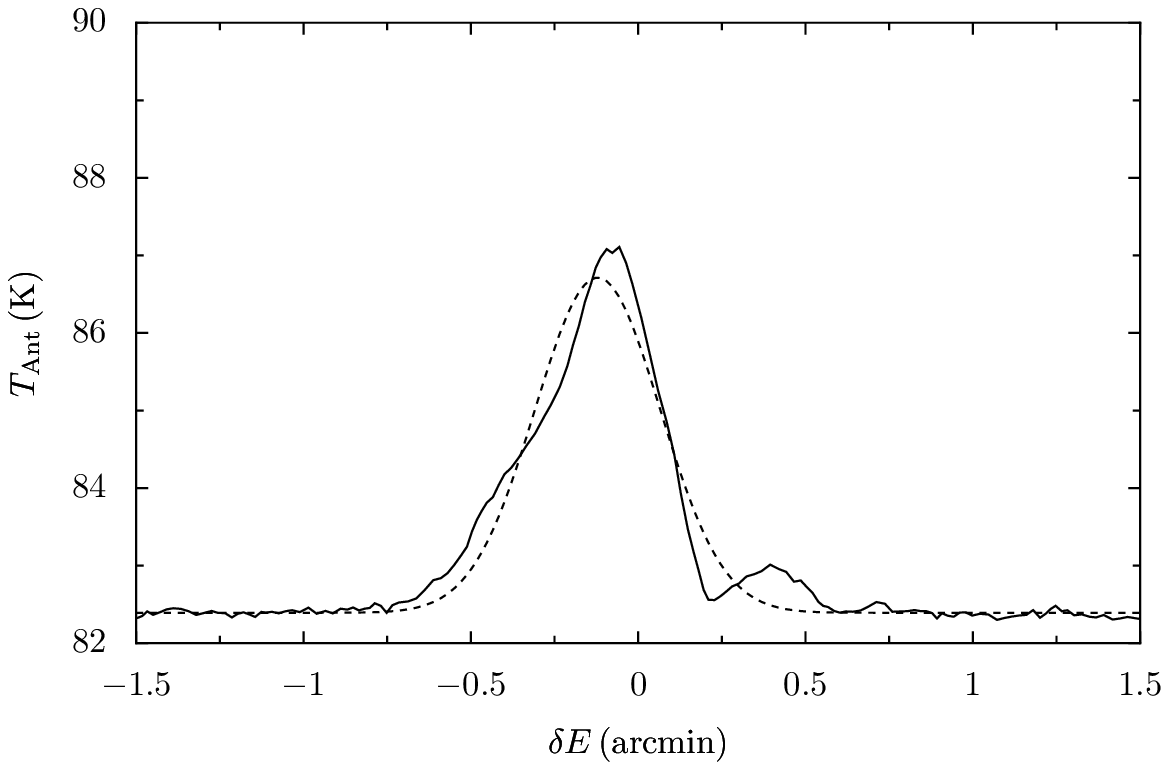}

  \includegraphics[width=0.99\columnwidth]{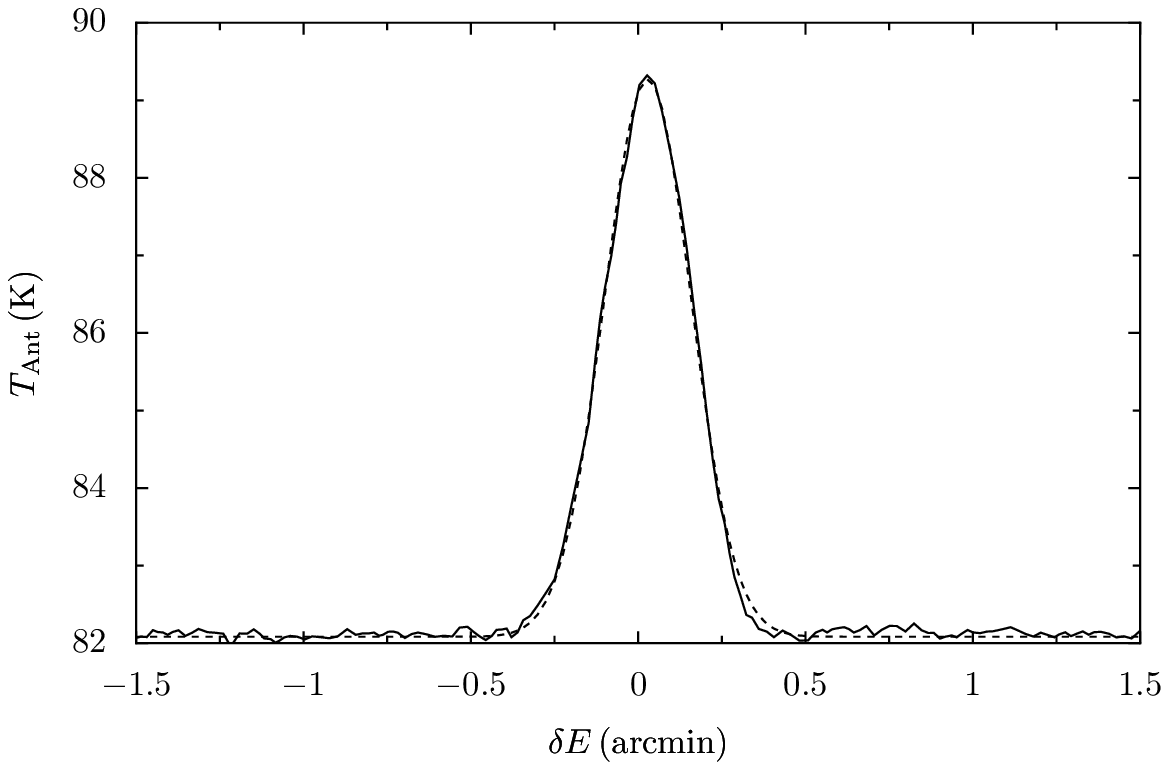}

  \caption{Elevation scans through a calibrator source before (top)
  and after (bottom) applying corrections derived from OOF
  measurements. Dashed lines show best-fitting Gaussian models to the
  scans.}

  \label{fig:gbtdayelscans}

\end{centering}
\end{figure}

Finally, as an independent, and for astronomical observing most
important, measure of the successful application of the technique we
have performed scans across calibration sources in the azimuth and
elevation directions before and after applying corrections from OOF
observations. For the daytime example given above, the results of such
elevation scans are shown in Figure~\ref{fig:gbtdayelscans}. The
improvement is dramatic: the peak antenna temperature (i.e., the
on-axis gain of the antenna) was increased by 53\,\% by making
adjustments from the OOF measurement.  The side-lobes clearly visible
in the scan without any adjustments (top of
Figure~\ref{fig:gbtdayelscans}) have disappeared.

\section{Gravitational deformations}
\label{sec:grav-deform}

One of the main advantages of the OOF holography technique is of
course that it allows measurements to be made over the full range of
elevations at which astronomical observations are normally carried
out.  This makes it possible to measure the aberrations which are due
to gravitational deformation of the telescope but are not fully
corrected for by the active surface and the standard finite-element
model.  Such aberrations can then be corrected by operating the active
surface using the sum of the FE model and an empirical model derived
from the measurements.

In order to measure these gravitationally-induced aberrations, we
carried out a campaign of 36 OOF observations over three nights in the
Spring and Autumn of 2005. The FE model was applied through the active
surface as normal for these observations and so we were measuring only
the residual deformation. In order to minimise the effect of thermal
gradients in the telescope structure, the observations were carried
out only between approximately midnight and sunrise of each night;
measurements during the day or early evening show that aberrations due
to thermal effects are comparable to or exceed that due to
uncompensated gravity deformation.  After analysis these data produced
36 sets of coefficients of Zernike polynomials describing the
wavefront errors over a wide range of elevations.

\begin{figure}
  \begin{centering}

  \includegraphics[width=0.99\columnwidth]{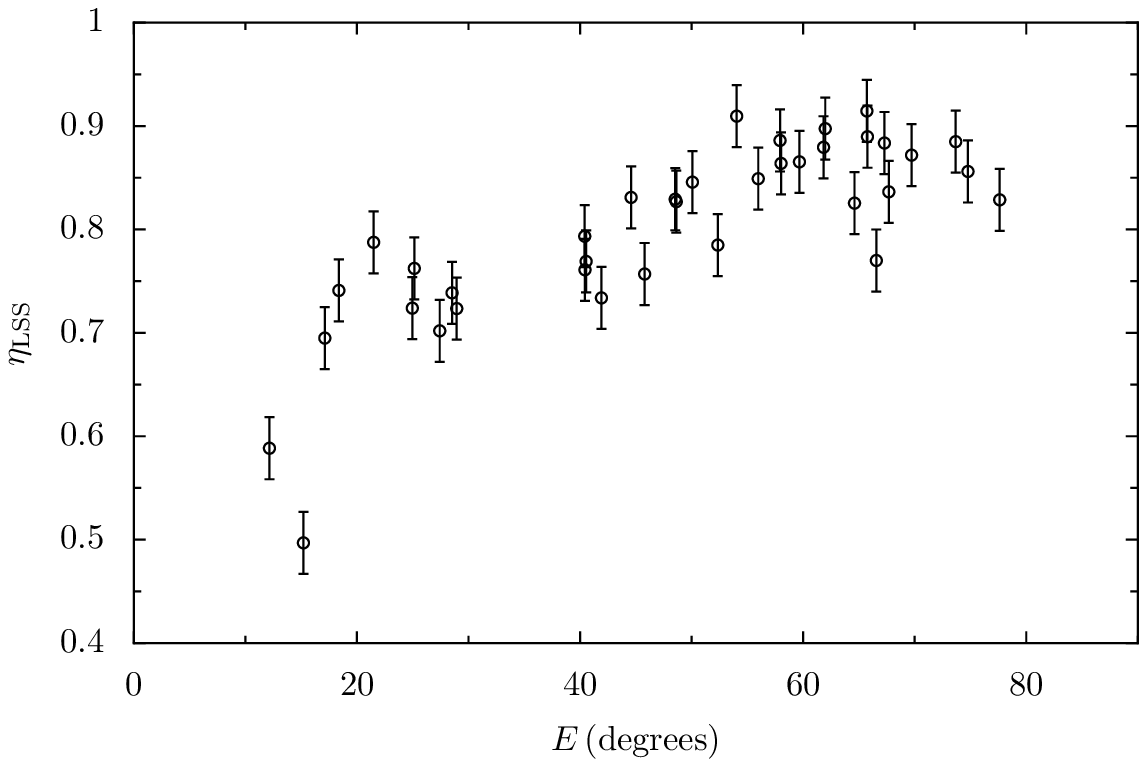}

  \caption{Predicted decrease of the on-axis gain at 43\,GHz due to
  measured large-scale structure ($\eta_{\rm LSS}$) derived from OOF
  measurements over a range of elevations ($E$).}

  \label{fig:offseteff}

  \end{centering}
\end{figure}

In order to examine if the OOF measurements support the hypothesis
that the existing FE model does not fully correct for gravitational
deformation of the GBT, we first computed the expected effect of the
large scale errors on the telescope gain. This is summarised in
Figure~\ref{fig:offseteff}, which shows the expected decrease in
on-axis gain due to the large-scale structure derived from each OOF
measurement plotted against elevation at which the observation was
made. The efficiency values have an estimated random error of 3\%,
corresponding to the 100\,\micron\ weighted RMS error of the wavefront
maps. It can be seen from the plot that the large-scale wavefront
errors level off around the rigging angle of 50.3 degrees but they
become increasingly important toward low elevations.  This is
consistent with gravitational deformation. It is interesting that we
measure a modest amount ($\approx150$\,\micron\ RMS) of large-scale
structure even at rigging angle, implying there is a systematic error
in the actuator zero-point settings. Correction of this large-scale
error produces a modest ($\approx10\,$\%) increase in efficiency at 43
GHz.

\begin{figure}

  \includegraphics[width=0.99\columnwidth]{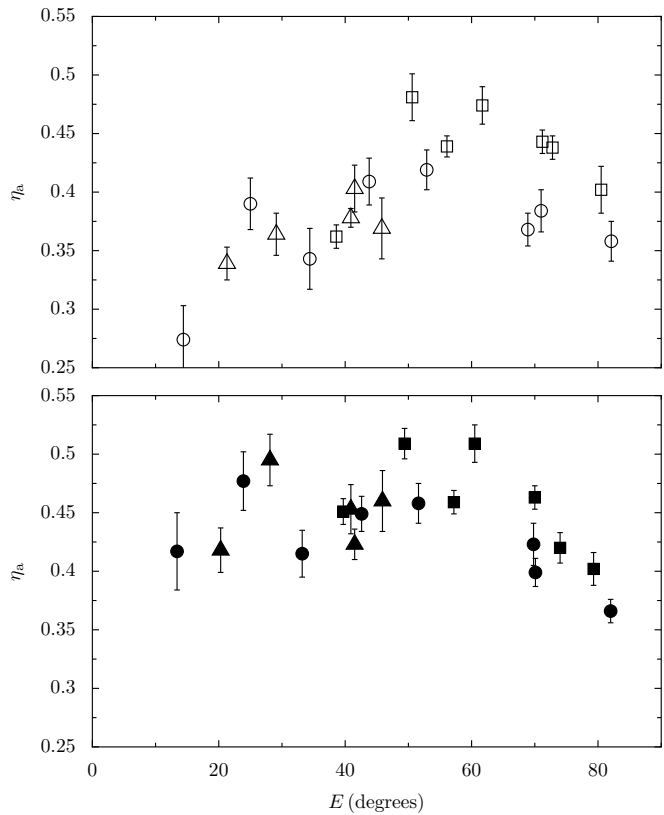}
  
  \caption{Aperture efficiency at 43\,GHz ($\eta_{\rm a}$)
  vs. elevation ($E$) measurements on 11 April 2005. Top panel, with
  open symbols, shows the measurements made with the nominal FE
  model; bottom panel, with filled symbols, shows the measurements
  with the FEM+OOF model applied. Sources used for the efficiency
  measurements are denoted by symbol types: 1642$+$3948 (square),
  1256$-$0547 (triangle) and 3C\,286 (circle). }
  \label{fig:msrgain1005}

\end{figure}

After the first night of OOF measurements, we made a simple model to
correct for these gravitational deformations by linearly interpolating
the coefficients as a function of elevation between the OOF
models. This model was tested by measuring the aperture efficiency of
the GBT as a function of elevation with only the FE model and the
finite-element plus correction derived from OOF measurement model
(FEM+OOF) applied.  The aperture efficiencies were measured by using
strong point sources which we bootstrapped from a primary flux
calibrator. These measurements were made in total power mode, by
interleaving measurements of empty sky and source, and using the same
frequency and receiver as used for holography. The results are shown
in Figure~\ref{fig:msrgain1005}.

Unfortunately, the aperture efficiency measurements suffer from
significant uncertainties due to fluctuations in atmospheric opacity,
as well as pointing and tracking errors of the telescope which are not
negligible when observing at 7\,mm. Nevertheless, it is quite clear
from the results shown that there is a measurable improvement to the
aperture efficiency at low elevation. Furthermore, when the FEM+OOF
model is applied, there does not appear to be significant remaining
variation of aperture efficiency with below 70\,degrees elevation. The
apparent fall-off at higher elevations may be due to effects not
associated with gravitational deformation, such as poorer tracking
performance of the telescope.

Using data from all three nights, we continued to investigate if the
shape of the measured deformations is consistent with the effects of
gravity.  As described above, the result of each individual
measurement is a set of coefficients of Zernike polynomials. Working
on the assumption that the coefficients are completely independent
\citep[this is only approximately true, see][]{OOFNikolic05p1} we
considered each one separately and found the best-fitting function of
the form expected for gravitational deformations.  If the telescope is
a linear elastic structure that is symmetric in the plane of the feed
arm, then gravitationally induced aberrations should depend on the two
resolved components of gravity only, that is, $g\sin (E)$ and $g\cos
(E)$, where $E$ is the elevation of the telescope.  Taking into
account the possibility of a further, gravitationally-independent,
setting error, the coefficients of Zernike polynomials, $z_{i}$,
should follow the functional form:
\begin{equation}
  z_{i}(E) = a \sin(E) + b \cos(E) + c.
  \label{eq:hookmodel}
\end{equation}

\begin{figure*}
  \includegraphics[width=0.3\linewidth]{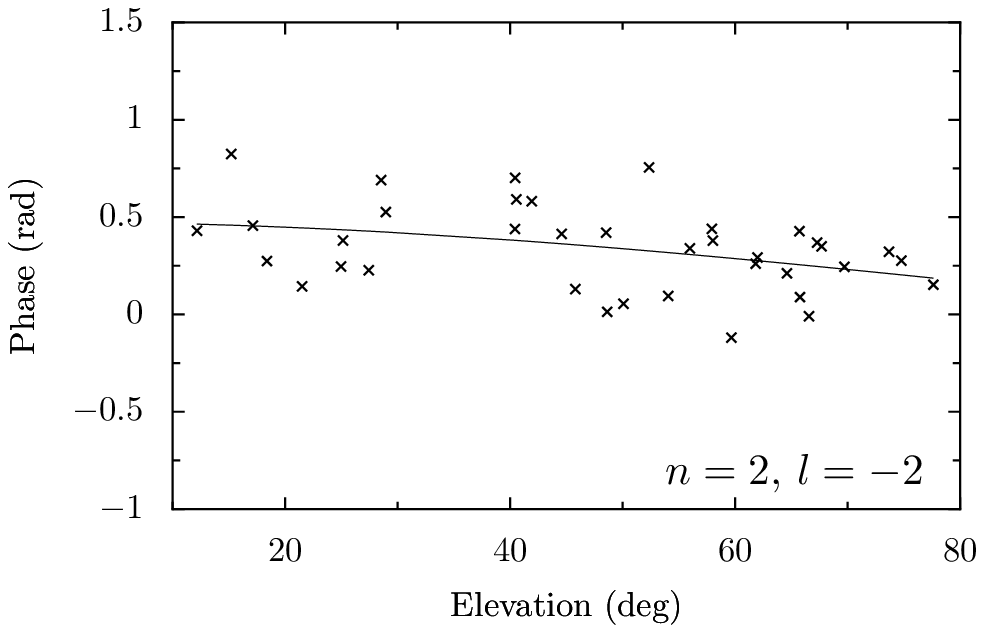}
  \includegraphics[width=0.3\linewidth]{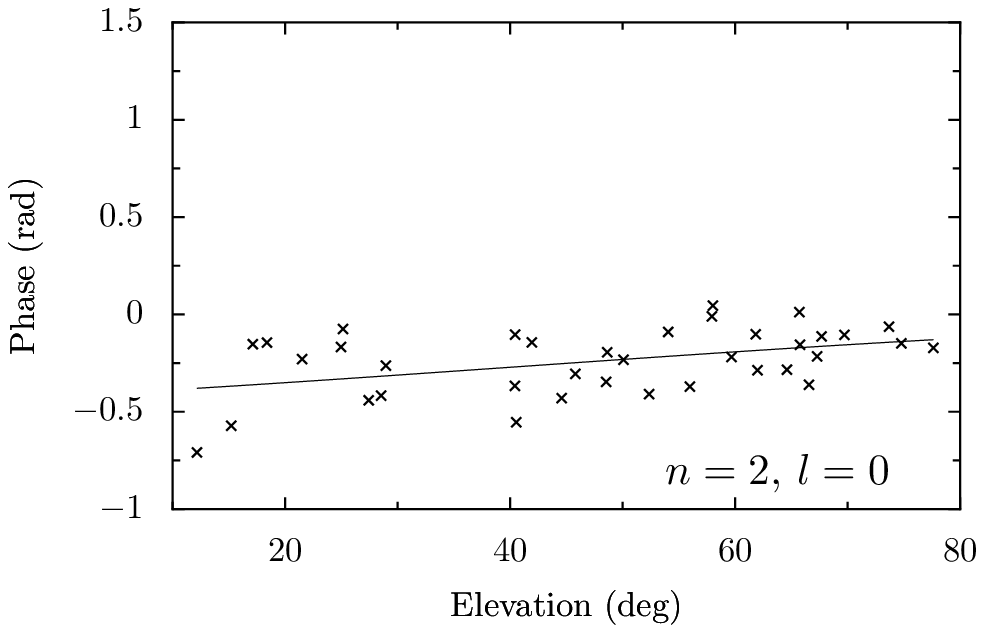}
  \includegraphics[width=0.3\linewidth]{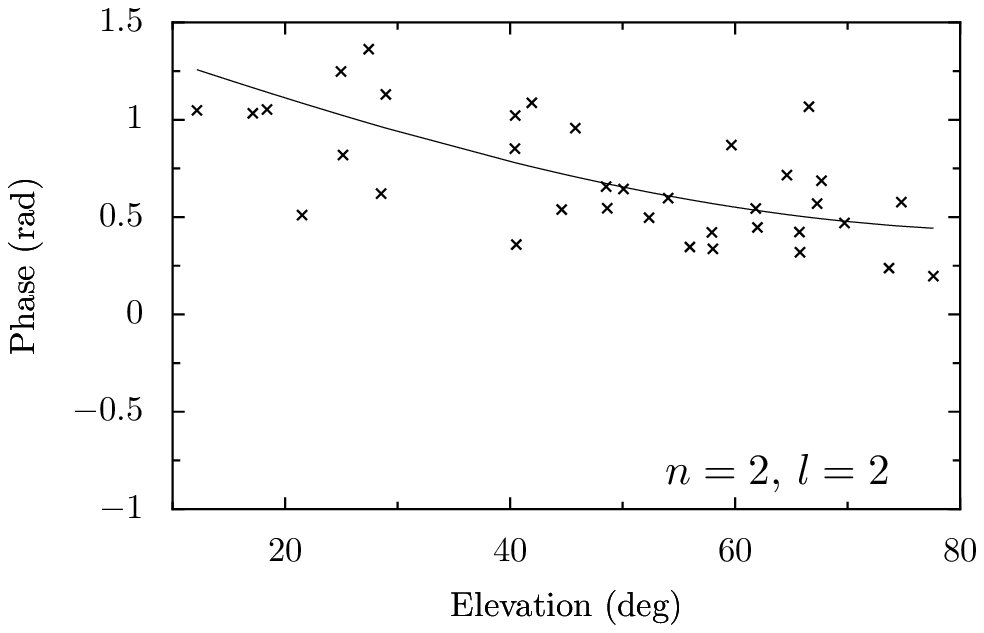}

  \includegraphics[width=0.3\linewidth]{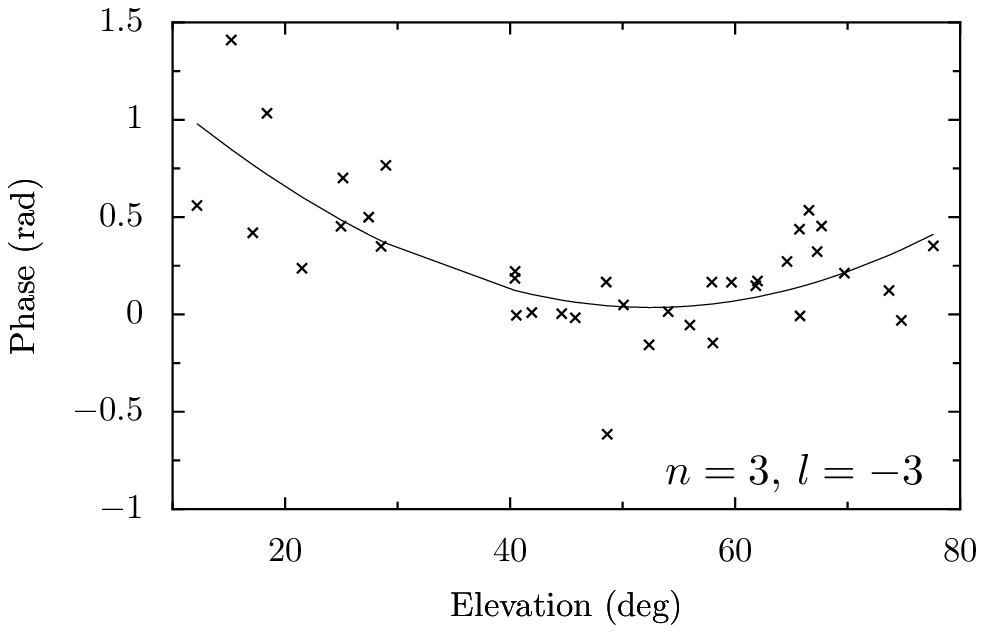}
  \includegraphics[width=0.3\linewidth]{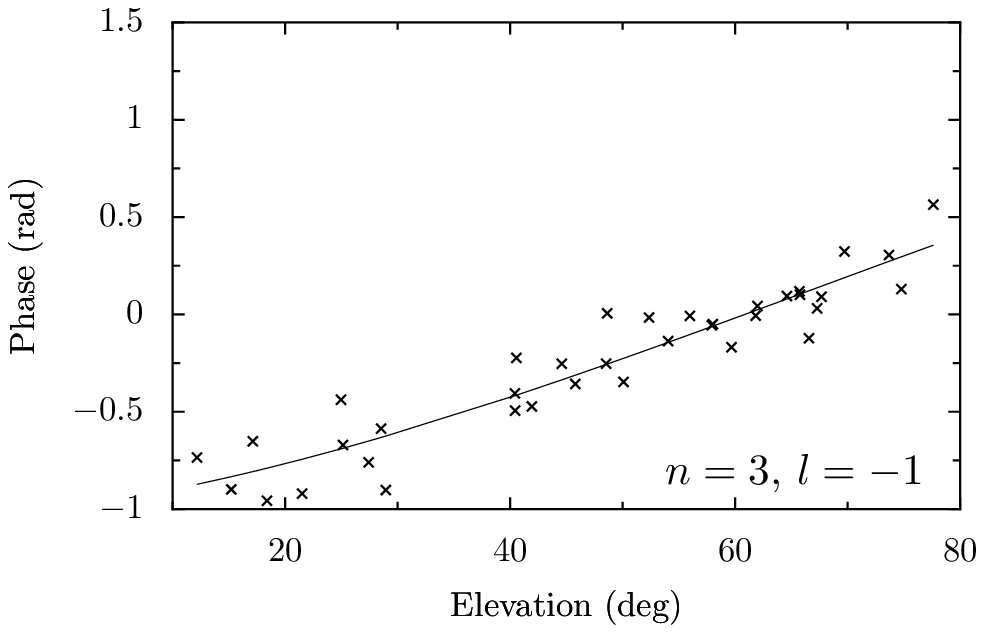}
  \includegraphics[width=0.3\linewidth]{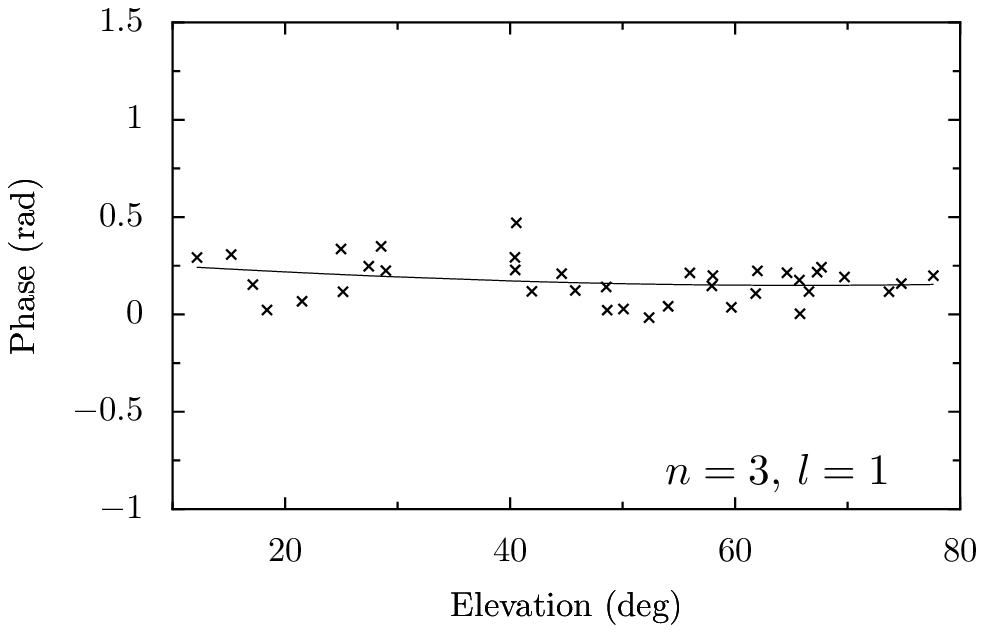}

  \includegraphics[width=0.3\linewidth]{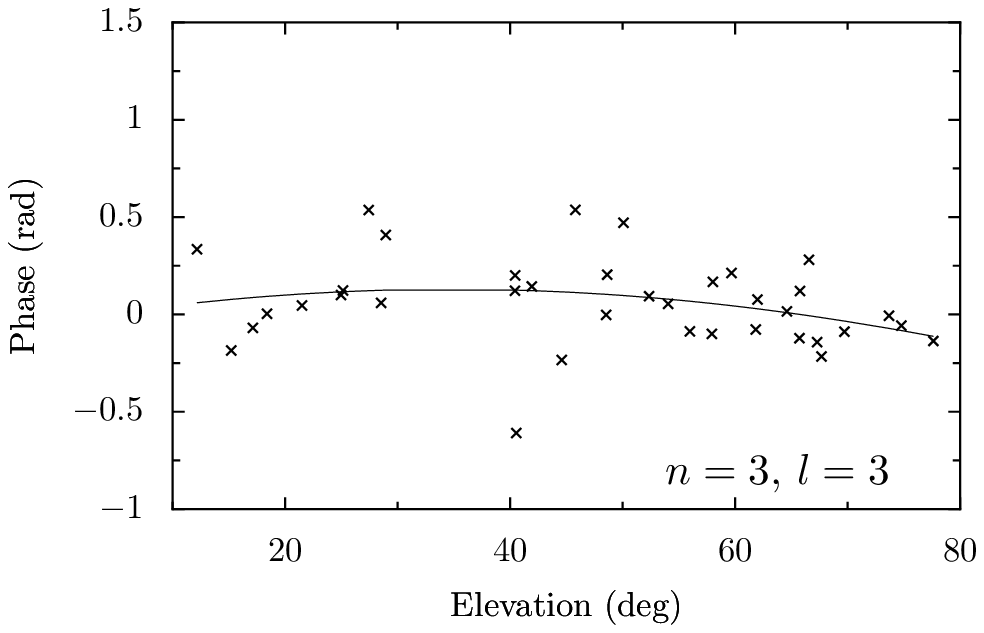}
  \includegraphics[width=0.3\linewidth]{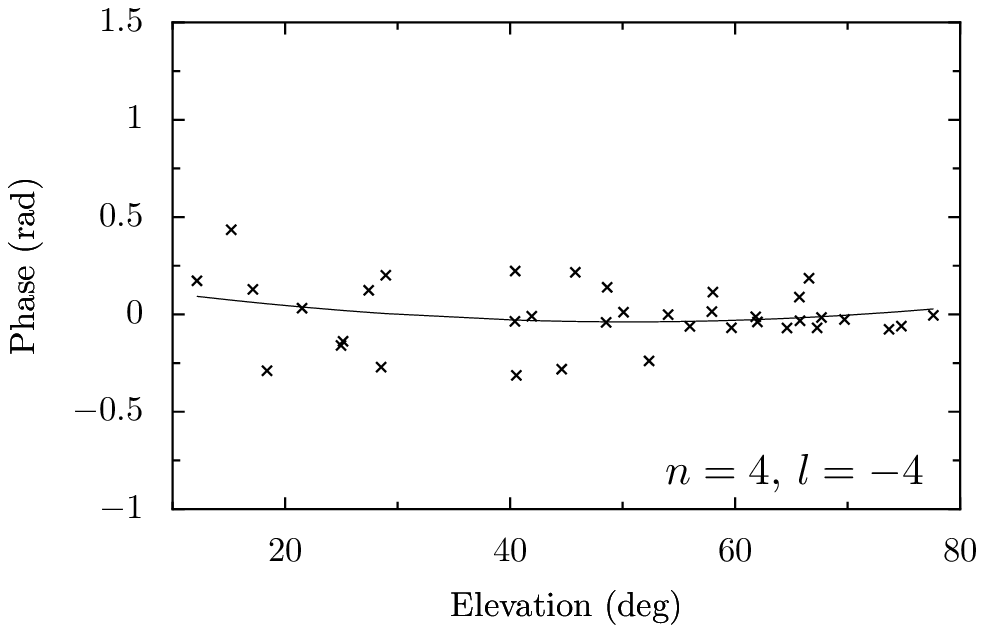}
  \includegraphics[width=0.3\linewidth]{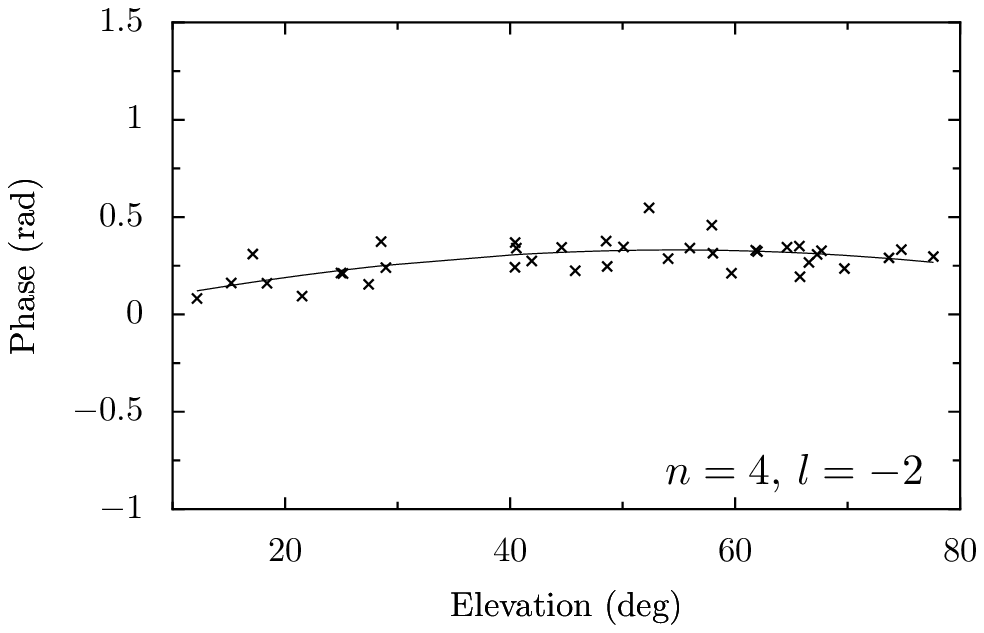}

  \includegraphics[width=0.3\linewidth]{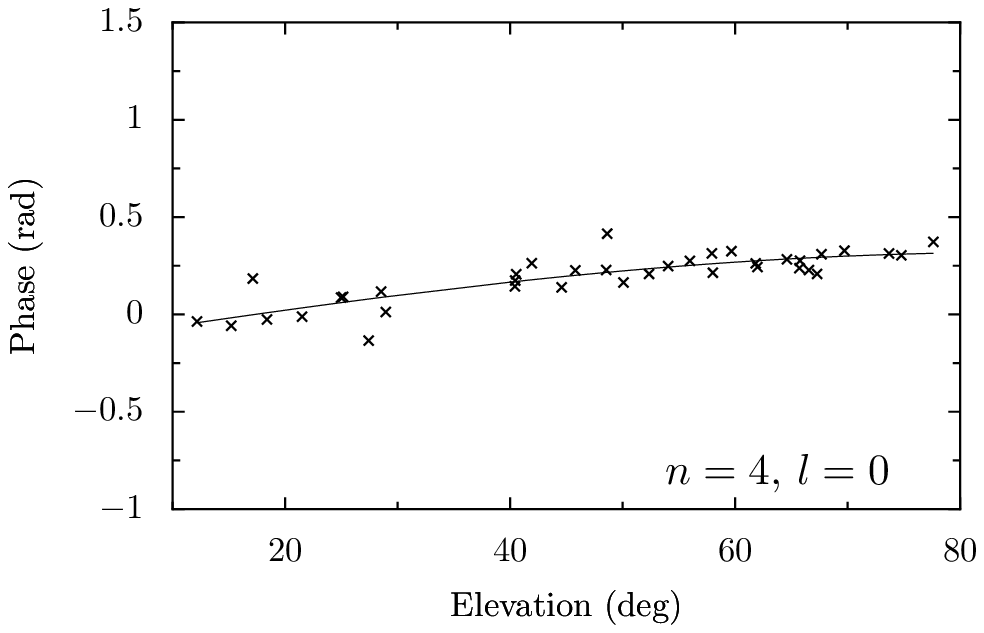}
  \includegraphics[width=0.3\linewidth]{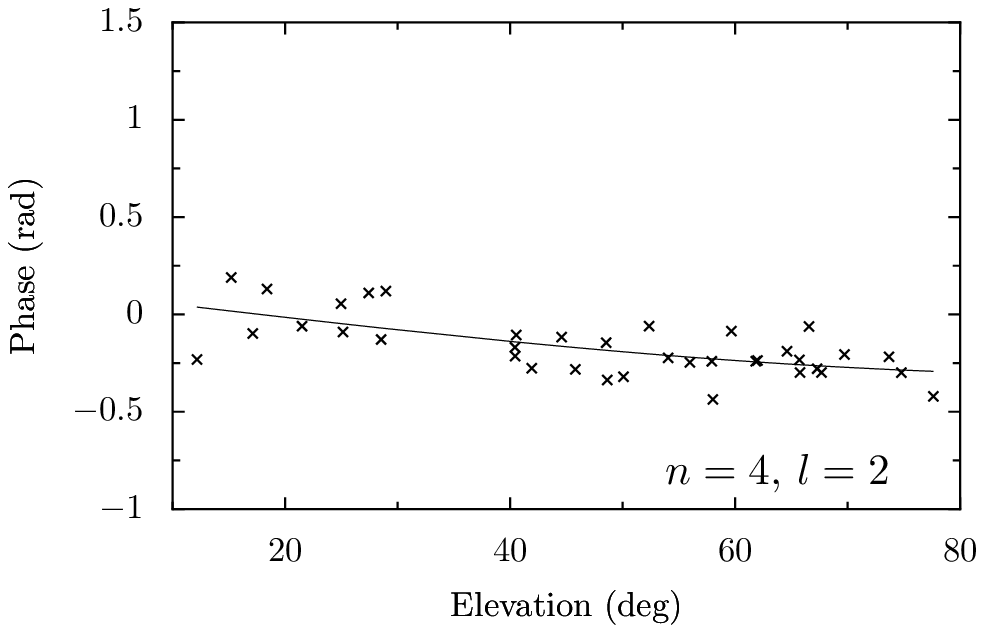}
  \includegraphics[width=0.3\linewidth]{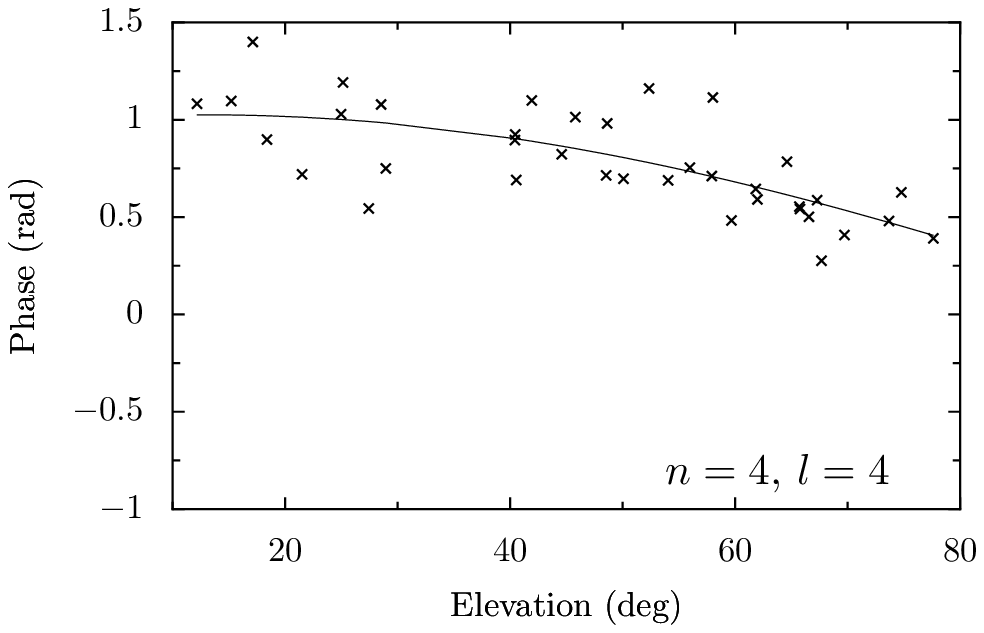}

  \includegraphics[width=0.3\linewidth]{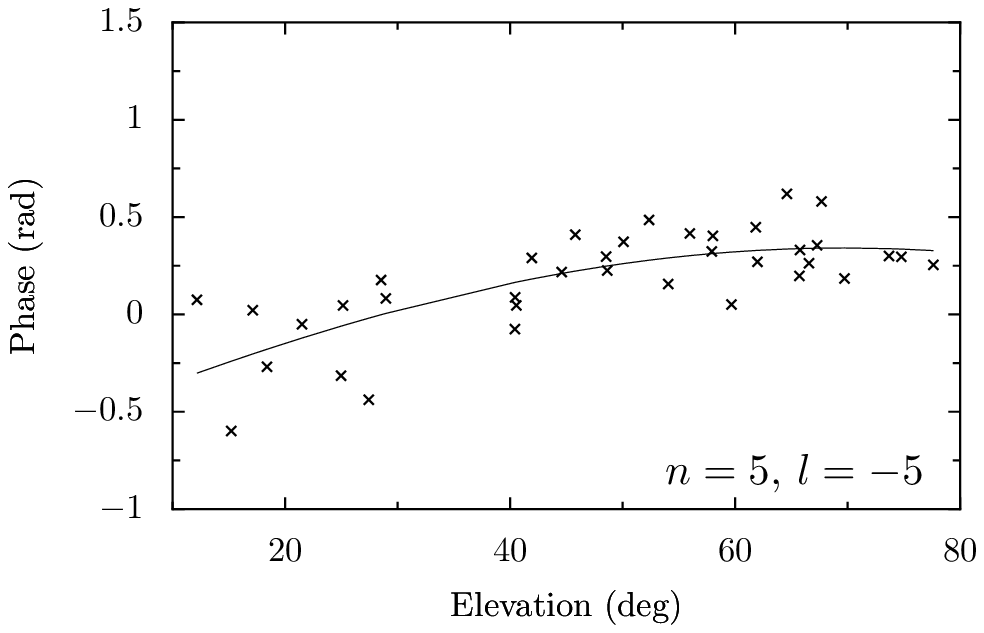}
  \includegraphics[width=0.3\linewidth]{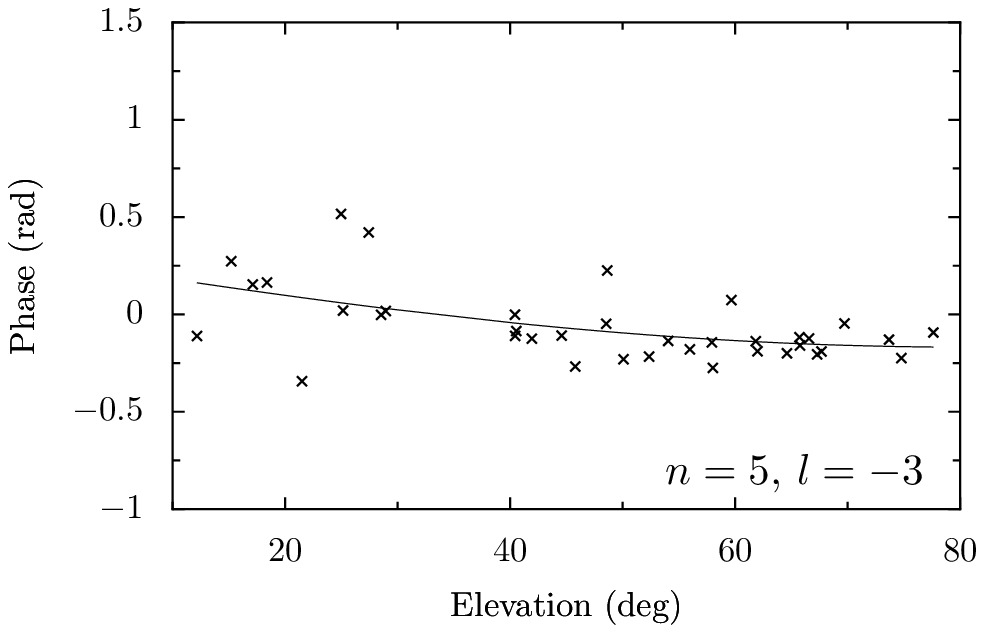}
  \includegraphics[width=0.3\linewidth]{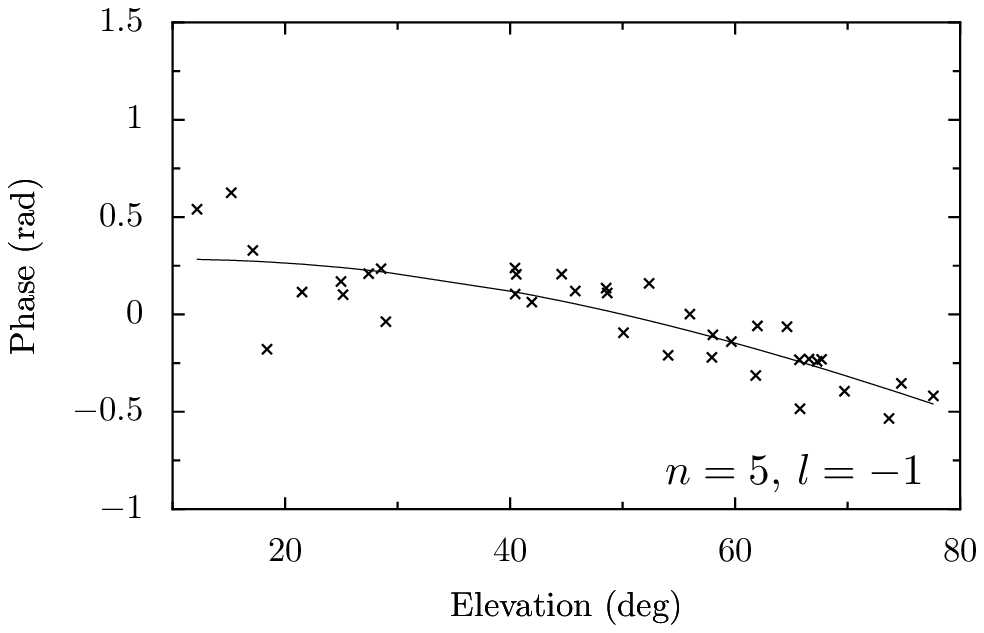}

  \includegraphics[width=0.3\linewidth]{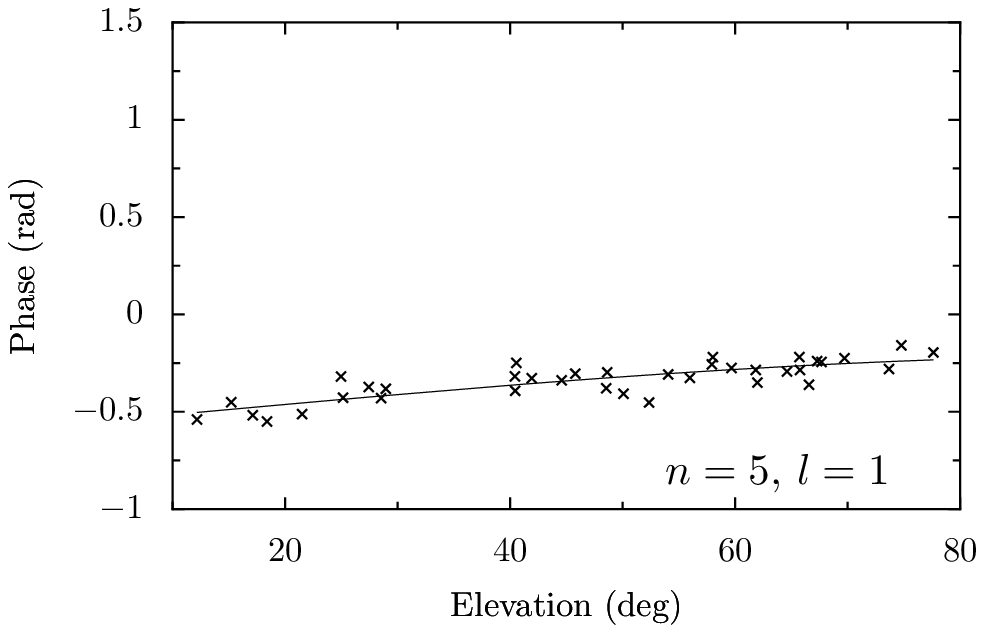}
  \includegraphics[width=0.3\linewidth]{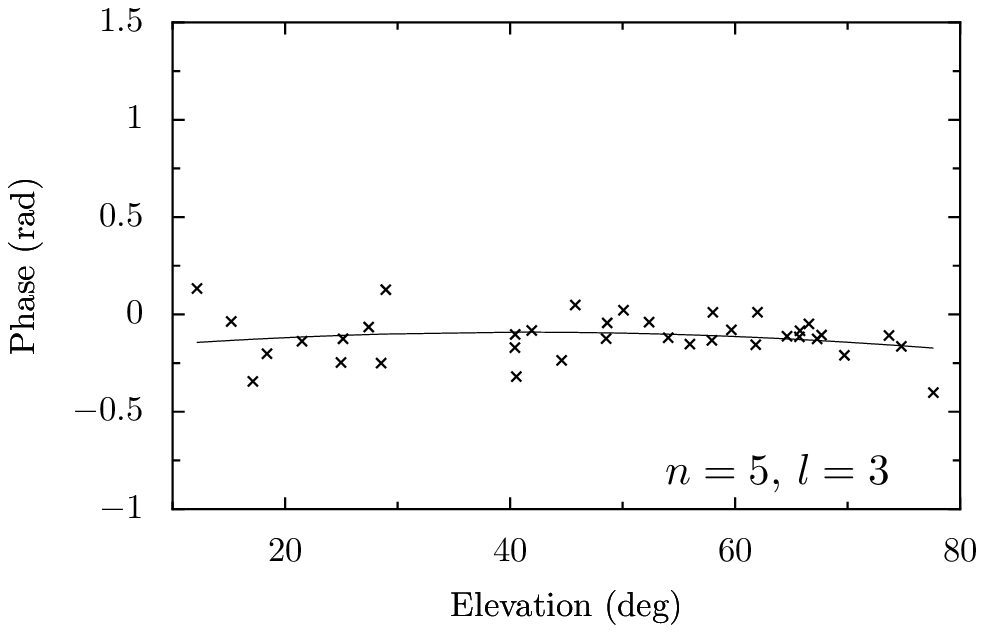}
  \includegraphics[width=0.3\linewidth]{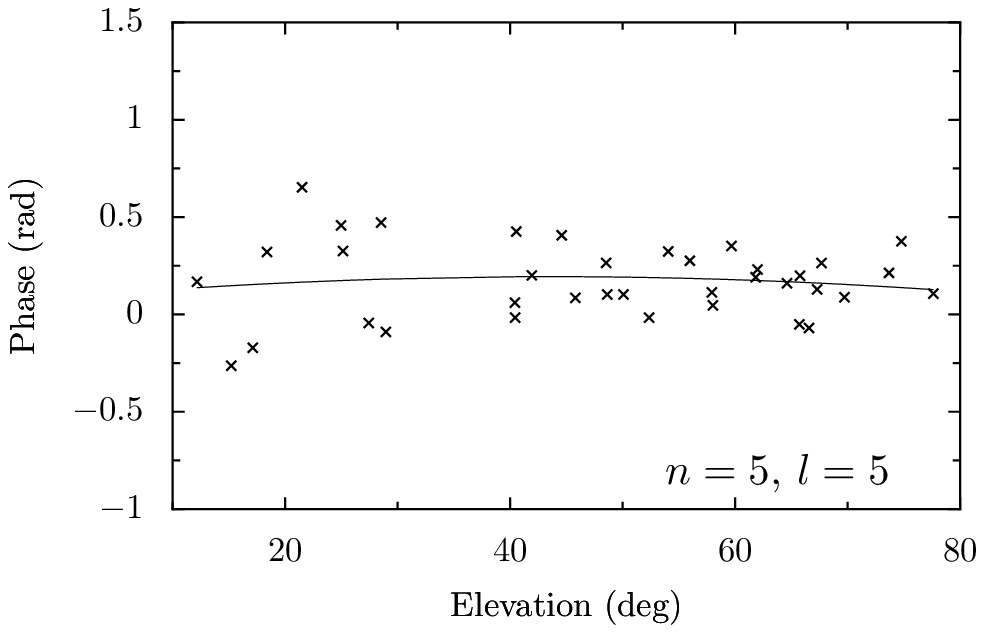}

  \caption{Measured values of the coefficients of Zernike polynomials
  as a function of elevation (crosses) and the best fitting model of
  the form shown in Equation~\ref{eq:hookmodel} (solid line).  Units
  throughout are radians of phase at the aperture edge (one radian of
  phase corresponds to a surface displacement 550\,\micron\ in the
  normal direction), $n$ is the radial and $l$ is the angular order of
  the polynomial.}
  \label{fig:gravmodfit}
\end{figure*}

This then was the function which was fitted to the observed
coefficients, with all $a$, $b$ and $c$ being free parameters. The
observed data and the best-fitting functions of the form in
Equation~\ref{eq:hookmodel} are shown in Figure
\ref{fig:gravmodfit}. As it can be seen in the figure, the measured
values of many of the coefficients have an elevation dependence that
is clearly consistent with gravitational deformation. The coefficients
which show little trend with elevation (e.g., $n=3$, $l=1$ and $n=3$,
$l=3$) tend to correspond to Zernike polynomials that are strongly
asymmetric in the cross-elevation direction.

Within the uncertainties caused by other effects, as expected the
gravitational deformations appear completely repeatable. The full
linear elastic structure model described above has been incorporated
into the GBT active surface control system, and is now in routine
operation.

\section{Discussion}

We begin by re-iterating that the OOF technique measures only
wavefront errors which are correlated over large distances in the
aperture plane. In this study we described surface errors as a linear
combination of 18 Zernike polynomials.  Larger beam maps with better
signal to noise ratios (as may be possible to obtain using focal plane
arrays) will provide opportunities to increase the size of the basis
set used in the fitting procedure, but it is unlikely it will be
possible to use this technique to set the 2000 individual panels of
the GBT.  The OOF technique is nevertheless useful because effects
such as gravitational and thermal deformations, and mis-alignment of
the optical elements, naturally induce large-scale wavefront errors.

In the first experiment, described in
Section~\ref{sec:retr-known-deform}, we demonstrate the ability of the
OOF technique to measure deformations of the telescope surface.  The
combination of relatively short measurement time and the GBT active
surface system allowed us to introduce a realistic sized-bump to the
primary reflector (the bump peak was around 700\,\micron\ or
$\lambda/10$) and the surface map retrieved using the OOF technique is
clearly very similar to the applied bump.  The illumination weighted
RMS error of the bump measurement was 130\,\micron.  As described in
Section~\ref{sec:closure}, when no large deformation is applied to the
surface, the random accuracy of the OOF technique can be as low as
70\,\micron\ (i.e., $\lambda/100$). If OOF measurements are made at
the shortest wavelengths at which astronomical observing are to be
carried out -- which should normally be possible because only the
standard astronomical receivers are used -- this accuracy should be
more than adequate for practical application.

We then investigated the practical applications of these measurements
of wavefront errors. Our main target was gravitational deformations
because, as they depend on elevation only they are expected to be
repeatable, and because the variation of the GBT gain with elevation
suggests that they are present even when the FE model is used.  We
used a series of OOF measurements to derive an elevation-dependent
correction to the FE model, and with these corrections applied to the
telescope, we find that the measured aperture efficiency is largerly
independent of elevation (filled symbols in
Figure~\ref{fig:msrgain1005}). This suggests that, as expected, the
major aberrations induced by gravity are large-scale and measurable
using the OOF technique, and, that they can be successfully corrected.
We note that such measurements and a model for gravitational
deformation may be useful even for telescopes without an active
surface. For example, if the panels of the primary surface are set
using holography with a ground-based transmitter, i.e., essentially at
zero degree elevation, it would normally be desirable to make an
adjustment to this setting so that the expected gravitational
deformations are as small as possible at higher elevation, close to
the elevations used for typical observations. Such an adjustment could
easily be determined using the technique described here.

We also investigated measurement and correction of aberrations due to
thermal effects on the telescope structure. These thermal deformations
of the GBT optics can easily dominate the total wavefront error during
the daytime as evidenced by degradation of the aperture efficiency and
beam shape of the GBT after sunrise.  In Section~\ref{sec:closure} we
showed that corrections derived from OOF measurements during the
daytime can dramatically improve both the gain and side-lobe structure
of the telescope. This demonstrates that the OOF technique described
here is fully capable of measuring and correcting for thermal effects.
Subsequent daytime measurements have confirmed, as expected, that the
deformations due to these effects are large-scale, and smoothly
varying with time. Unfortunately, although thermal effects are
expected to be fully repeatable in the sense that the same thermal
environment should produce the same deformation; the varying thermal
conditions, and orientation history of the telescope from day to day
means that there is no single parameter which can be used to
parameterise the deformations, analagous to the telescope elevation
for the gravitational deformations.

The time it currently takes to obtain an OOF data-set with our
dual-pixel receiver means that it is impractical either to attempt to
characterise the entire range of thermal environments with OOF maps,
or to perform routine measurements and adjustments in ``real time''.
The results shown are, however, a proof of principle that one of these
approaches (most likely the real time correction) will be useful once
multi-pixel focal-plane arrays become available.  With focal-plane
arrays, the time required to make the beam maps will not be much
longer than required for optimising the focus of the telescope. In
this case, the OOF analysis used here may be used to correct the focus
\emph{and} higher order aberrations whenever a traditional focusing
measurement would have been done.

\section{Summary}

We have performed ``OOF'' phase-retrieval holography measurements of the
100-metre Green Bank Telescope, using the technique described by
\cite{OOFNikolic05p1}. The advantages of the OOF technique which
motivated us to carry out these observations are:
\begin{enumerate}
  \item The measurements provide a map of total large-scale wavefront
  errors (aberrations) present in the system.
  \item The aberrations can be measured over the full range of
  elevations making it possible to infer the effects of gravity on the
  telescope.
  \item The time required for each measurement is relatively short.
  \item The covenience of being able to perform the measurements using 
  the standard astronomical receivers, with no special hardware
  or software setup, brings considerable operational advantages. 
\end{enumerate}
In addition, and very importantly, the active surface of the GBT has
allowed us to immediately and easily test our measurements.

As far as general practicalities of using the OOF technique are
concerned: 
\begin{enumerate}
  \item We have demonstrated that OOF holography with astronomical
  sources and receivers can be routinely carried out on a large
  millimeter-wave radio telescope. A full set of observations takes of
  the order of 25 minutes.
  \item Observing at a wavelength of 7\,mm, using a
  measure-correct-measure cycle, we measure an illumination-weighted
  half-path large-scale error of around 100\,\micron, indicating that
  the \emph{random} error associated with the OOF technique in this
  case is $\approx \lambda/100 $.
  \item Correcting for the measured aberrations produces an
  improvement in the aperture efficiency close to the expected
  improvement.
\end{enumerate}

More specifically to the GBT, we found:
\begin{enumerate}
  \item At night time and at elevations close to the rigging angle,
  there is a small but measurable large-scale wavefront error.  
  \item By making OOF maps at a number of elevations we have been able
  to derive a model for aberrations which is consistent with
  gravitational deformations. This model, which is essentially a small
  refinement to the existing finite element model for the telescope,
  is now in routine use for high-frequency observing.
  \item Applying this model makes an improvement to the aperture
  efficiency at elevations smaller than 40 degrees, largely making the
  aperture efficiency independent of elevation. This indicates that
  the deformations induced by gravity are large-scale and measurable
  using this technique.
  \item Daytime thermal deformations of the GBT are also large scale,
  and smoothly varying with time. The time currently required to obtain an OOF
  dataset makes attempts to correct for
  thermal deformations impractical at this time. However, this technique
  should become directly applicable once array receviers are available
  on the telescope.

\end{enumerate}

\begin{acknowledgements}

We would like to thank the referee for his comments which
significantly improved this paper.

The National Radio Astronomy Observatory is a facility of the National
Science Foundation operated under cooperative agreement by Associated
Universities, Inc. B.~Nikolic would like to thank NRAO for the
appointment to a post-doctoral position during the course of which the
work described here was carried out. We acknowledge George
Moellenbroack for assistance with data pre-processing software in the
early stages of this project.

Although we have carried out the investigations described here, the
GBT active surface itself, and the associated control software, is the
result of many years of hard work by numerous past and present members
of the NRAO staff. We applaud them for providing us with such a
wonderful system to work with.

\end{acknowledgements}

\bibliographystyle{aa} 

\bibliography{oofbib.bib}

\end{document}